\documentclass[iop,apj]{emulateapj}

\usepackage{natbib}
\usepackage{amsmath}
\usepackage{multirow}
\usepackage{float}
\usepackage{color}
\usepackage{afterpage}
\usepackage[abs]{overpic}
\usepackage{CJK}
\usepackage{hyperref}
\usepackage{rotating}

\definecolor{Red}{rgb}{1,0,0}
\definecolor{Blue}{rgb}{0,0,1}
\definecolor{Green}{rgb}{0,1,0}
\definecolor{magenta}{rgb}{1,0,.6}
\definecolor{lightblue}{rgb}{0,.5,1}
\definecolor{lightpurple}{rgb}{.6,.4,1}
\definecolor{gold}{rgb}{.6,.5,0}
\definecolor{orange}{rgb}{1,0.4,0}
\definecolor{hotpink}{rgb}{1,0,0.5}
\definecolor{newcolor2}{rgb}{.5,.3,.5}
\definecolor{newcolor}{rgb}{0,.3,1}
\definecolor{newcolor3}{rgb}{1,0,.35}
\definecolor{darkgreen1}{rgb}{0, .35, 0}
\definecolor{darkgreen}{rgb}{0, .6, 0}
\definecolor{darkred}{rgb}{.75,0,0}

\def \arcmin     { ^{\prime} }
\def \arcsec    {^{\prime\prime}}

\def \kms            {~{\rm km~s}^{-1}}

\def \etal      {et~al.\ }

\def \hmsol     {h^{-1}{\rm\ M}_\odot}
\def \hMpc      {h^{-1}{\rm\ Mpc}}

\def \h         {\hbox{$\, h$} }

\newcommand{\nhi}{\ensuremath{N_\mathrm{HI}}}
\newcommand{\persqcm}{\,\mathrm{cm^{-2}}}
\newcommand{\snr}{\ensuremath{\mathrm{S/N}}}
\newcommand{\lya}{Ly$\alpha$}

\newcommand{\lyaf}{Ly$\alpha$ forest}
\newcommand{\waveion}[3]{\ion{#1}{#2} $\lambda$#3}

\newcommand{\scien}[2]{#1  \times 10^{#2}} 
\newcommand{\beq}{\begin{equation}}
\newcommand{\eeq}{\end{equation}}
\newcommand{\bc}{\begin{center}}
\newcommand{\ec}{\end{center}}
\newcommand{\bfig}{\begin{figure}}
\newcommand{\efig}{\end{figure}}

\newcommand{\zalp}{\ensuremath{z_\alpha}}

\newcommand{\ang}{\ensuremath{\mathrm{\AA}}}

\newcommand{\dperp}{\ensuremath{\langle d_{\perp} \rangle}}
\newcommand{\sigthreed}{\ensuremath{\epsilon_{\rm 3D}}}

\newcommand{\delrecon}{\ensuremath{\delta_F^{\rm rec}}}
\newcommand{\delsm}{\ensuremath{\delta_F^{\rm sm}}}
\newcommand{\deldm}{\ensuremath{\delta_\mathrm{dm4}}}
\newcommand{\sigsm}{\ensuremath{\sigma_{\rm sm}}}

\newcommand{\persqdeg}{\mathrm{deg}^{-2}}
\newcommand{\sqdeg}{\mathrm{deg}^{2}}
\newcommand{\cmd}{\mathbf{C}_\mathrm{MD}}
\newcommand{\cdd}{\mathbf{C}_\mathrm{DD}}

\newcommand{\yperp}{\ensuremath{y_\mathrm{perp}}}

\newcommand{\Lmax}{\ensuremath{L_{\rm max}}}

\def \etal {et~al.}

\slugcomment{First draft; To be submitted to ApJ}
\shorttitle{IGM Tomography of $z=2.45$ Protocluster in COSMOS}
\shortauthors{Lee \etal}

\begin{document}

\title{Shadow of a Colossus: A
$\lowercase{z} = 2.45$ Galaxy Protocluster Detected\\  in 3D \lya\ Forest Tomographic Mapping of the COSMOS Field}
\author{Khee-Gan Lee\altaffilmark{1}, 
Joseph F. Hennawi\altaffilmark{1},
Martin White\altaffilmark{2,3}, J.\ Xavier Prochaska\altaffilmark{4,5},  Andreu Font-Ribera\altaffilmark{3}, \\
David J.\ Schlegel\altaffilmark{3},  R.\ Michael Rich\altaffilmark{6}, Nao Suzuki\altaffilmark{7},  Casey W.\ Stark\altaffilmark{2,3}, 
 Olivier Le F\`evre\altaffilmark{8}, \\
Peter E.\ Nugent\altaffilmark{3},
Mara Salvato\altaffilmark{9}
Gianni Zamorani\altaffilmark{10},
%add zamorani, le-fevre, salvato
}
\altaffiltext{1}{Max Planck Institute for Astronomy, K\"{o}nigstuhl 17, D-69117 Heidelberg, West Germany}
\altaffiltext{2}{Department of Astronomy, University of California at Berkeley, B-20 Hearst Field Annex \# 3411,
Berkeley, CA 94720, USA}
\altaffiltext{3}{Lawrence Berkeley National Laboratory, 1 Cyclotron Rd., Berkeley, CA 94720, USA}
\altaffiltext{4}{Department of Astronomy and Astrophysics, University of California at Santa Cruz, 1156 High Street, Santa Cruz, CA 95064, USA}
\altaffiltext{5}{University of California Observatories, Lick Observatory, 1156 High Street, Santa Cruz, CA 95064, USA}
\altaffiltext{6}{Department of Physics and Astronomy, University of California at Los Angeles, Los Angeles, CA 90095-1562, USA}
\altaffiltext{7}{Kavli Institute for the Physics and Mathematics of the Universe (IPMU), The University of Tokyo, 
Kashiwano-ha 5-1-5, Kashiwa-shi, Chiba, Japan}
\altaffiltext{8}{Aix Marseille Universit\'e, CNRS, LAM (Laboratoire d'Astrophysique  de Marseille) UMR 7326, 13388, Marseille, France}
\altaffiltext{9}{Max Planck Institute for Extraterrestrial Physics, Gie{\ss}enbachstra§e 1, 85741 Garching bei M\"unchen, Germany}
\altaffiltext{10}{INAF--Osservatorio Astronomico di Bologna, via Ranzani,1, I-40127, Bologna, Italy}
\email{lee@mpia.de}

\begin{abstract}
Using moderate-resolution optical spectra from 58 background Lyman-break galaxies and quasars at $z\sim 2.3-3$ within a $11.5\arcmin\times13.5\arcmin$ area of the COSMOS field ($\sim 1200\,\persqdeg$ projected area density or $\sim 2.4\,\hMpc$ mean 
transverse separation), 
we reconstruct a 3D tomographic map of the foreground \lyaf\ absorption at
$2.2<z<2.5$ with an effective smoothing scale of $\sigthreed\approx3.5\,\hMpc$ comoving.
Comparing with 61 coeval galaxies with spectroscopic redshifts in the same volume,
we find that the galaxy positions are clearly biased towards regions with enhanced IGM absorption
in the tomographic map.
We find an extended IGM overdensity with deep absorption troughs 
at $z=2.45$
associated with a recently-discovered galaxy protocluster at the same redshift. 
Based on simulations matched to our data, 
we estimate the enclosed dark matter mass within this IGM
overdensity to be $M_{\rm dm} (z=2.45) = \scien{(9\pm4)}{13}\,\hmsol$, and argue based on this mass and absorption strength
that it will form at least one $z\sim0$ galaxy cluster with $M(z=0) = \scien{(3\pm2)}{14}\,\hmsol$,
although its elongated nature suggests that it will likely collapse into two separate clusters.
We also point out a compact overdensity of six MOSDEF galaxies at $z=2.30$ within a $r\sim 1\,\hMpc$ radius
and $\Delta z\sim 0.006$, 
which does not appear to have a large associated IGM overdensity.
These results demonstrate the potential of \lya\ forest tomography on larger volumes 
to study galaxy
properties as a function of environment, as well as revealing the large-scale IGM overdensities associated
with protoclusters or other features of large-scale structure.
\end{abstract}


\section{Introduction}

The study of high-redshift ($z>2$) galaxy protoclusters is a topic of increasing interest, providing a route to studying galaxy evolution, AGN, plasma physics, and our models of gravity.
They are also critical laboratories for understanding the growth of massive galaxy clusters at $z\sim0$.
While many studies have successfully found protoclusters around active supermassive black holes such as radio galaxies or luminous quasars (see \citealt{hennawi:2015} and \citealt{cooke:2014} for recent examples, or \citealt{chiang:2013} for a compilation),
it is difficult to tell whether these `signpost' protoclusters are representative of the overall population -- which, at least in simulations, shows a wide range of properties.

More uniform, `blind', search techniques are therefore required for unbiased samples of protoclusters. Several such objects have been found serendipitously in photometric or galaxy redshift surveys 
\citep[e.g.,][]{steidel:2005,gobat:2011,cucciati:2014,yuan:2014,casey:2015}.
Searching photometric  redshift catalogs \citep{chiang:2014} is another promising new
technique, but is limited to fields with extensive multi-wavelength photometry that enable accurate photometric redshfits.
Unfortunately we expect protoclusters to be rare, and surveying large volumes of space with these techniques is expensive.

Recently, \citet{stark:2015} (hereafter S15) argued that 3D tomographic reconstructions of the 
intergalactic medium (IGM) Lyman-$\alpha$ (\lya) forest absorption \citep{pichon:2001, caucci:2008, lee:2014} can be used to efficiently search for protoclusters.
A protocluster's \lya\ absorption signature extends over a large region ($r\gtrsim5\,\hMpc$) allowing it to be mapped by background galaxies separated by several transverse Mpc, corresponding to relatively bright limiting magnitudes ($\lesssim24.5\,$mag) and 
therefore accessible to existing 8-10m class telescopes.
In fact several studies have already found \lya\ absorption associated with protoclusters 
either in single sightlines \citep{hennawi:2015} or
by stacking multiple background spectra (\citealt{cucciati:2014}; Hayashino \etal\ in prep.).
Moderate-resolution spectra with careful sightline selection should enable 3D mapping even with modest signal-to-noise ratios (S/N).
 
In \citet{lee:2014a} (hereafter L14b), we made the first attempt at \lyaf\ tomography using a set of 24 background Lyman-break galaxy (LBG) spectra within a $5\arcmin\times12\arcmin$ area in the COSMOS field.
This marked the first-ever use of LBGs, instead of quasars, for \lyaf\ analysis, and the resulting
reconstruction at $2.20<z<2.45$ was the first 3D map of the $z>2$ universe probing Mpc-scale structures.
In this paper, as part of the pilot observations for the COSMOS Lyman-Alpha Mapping And Tomography Observations 
(CLAMATO) survey to map the high-redshift IGM, we extend the L14b map volume by a factor of $\sim2.5\times$.  This volume starts to approach that where we expect to see protocluster candidates in a blind
survey: in N-body simulations at $z\sim 2.5$ the comoving
number density of halo progenitors which by $z=0$ have masses above $10^{14}\,h^{-1}M_\odot$ is
$\sim 10^{-5}\,h^3\,{\rm Mpc}^{-3}$, or
to $10^{-6}\,h^3\,{\rm Mpc}^{-3}$ for the
progenitors of $3\times 10^{14}\,h^{-1}M_\odot$ halos \citep[e.g.][]{stark:2015}.
In the simulations of S15, mock observations with similar sightline density and noise as CLAMATO found more than 3/4 of the protoclusters which would eventually grow to form clusters more massive than $3\times 10^{14}\,h^{-1}M_\odot$ at $z=0$ but only 1/5 of the protoclusters which would form $10^{14}\,h^{-1}M_\odot$ halos by $z=0$.
Our observations probe a volume of $V=5.8\times 10^4\,h^{-1}{\rm Mpc}^3$ at $2.2<z<2.5$.
To the extent that these simulations are accurate, there
is a $\sim 10\%$ chance we should find a protocluster in the surveyed volume (see also \S\ref{sec:proto}).

Two other developments in the past year lend additional synergy: (i) the first data release of
the MOSDEF survey \citep[][hereafter K15]{kriek:2015}, whose near-infrared (NIR) nebular line redshifts reduce systemic redshift uncertainties hence allowing a cleaner comparison with our tomographic map; and
(ii) the discovery of a $z=2.45$ galaxy protocluster within our target field (\citealt{diener:2015}, hereafter D15; \citealt{chiang:2015}, hereafter C15).

As we shall see, our tomographic map suggests a complex 3D multi-pronged structure for this protocluster, while also revealing other intriguing structures.
We also find that galaxies co-eval with our map preferentially inhabit higher absorption regions of the IGM, suggesting that \lya\ tomography does indeed probe the underlying large scale structure.
 
This paper is organized as follows: we describe the data in
Section~\ref{sec:obs} and tomographic reconstruction of the foreground
IGM in Section~\ref{sec:tomo}, and then describe simulations used for
the subsequent analysis
(Section~\ref{sec:sims}). Section~\ref{sec:proto} then describes the
protoclusters and overdensities found within our map volume.
 
In this paper, we assume a concordance flat $\Lambda$CDM cosmology, with $\Omega_M=0.27$, 
$\Omega_\Lambda=0.73$ and $H_0 = 70\,\kms\,\mathrm{Mpc}^{-1}$.

\section{Observations}\label{sec:obs}

We targeted $2.3<z<3$ galaxies and AGN
 within the COSMOS field \citep{scoville:2007,capak:2007} in order to probe the foreground IGM
\lyaf\ absorption at $z\sim 2.3$. 
Top priority was given to objects with confirmed redshifts from the
zCOSMOS-Deep \citep{lilly:2007} and VUDS \citep{le-fevre:2015} spectroscopic surveys,
while we also, where available, added grism redshift information kindly provided by
the 3D-HST team \citep[e.g.][]{brammer:2012}. 
We also selected targets based on photometric redshifts from \citet{ilbert:2009} as well as
\citet{salvato:2011} for X-ray detected sources, which have $\sigma_z/(1+z)\approx 0.03$ and $<10\%$ catastropic failure rate. 
This target selection process was carried out in February 2014, prior to the publications by 
\citet{diener:2015} and \citet{chiang:2015} reporting galaxy overdensities in the field.

%From this overall catalog, we designed slitmasks aimed at ensuring spatial uniformity,
 %for background sources with redshift $\zbg$ probing \lya\ absorption
 %at redshifts\footnote{We assume that the \lya\ forest is within the restframe wavelength range
%$1040\,\ang < \lambda <1195\ang$ 
%of each background source, therefore for a given $\zalp$, the background sources are at
%$ (\lambda_{\alpha}/1195\,\ang)(1+\zalp)  <1+\zbg< (\lambda_{\alpha}/1040\,\ang)(1+\zalp)$, 
%where $\lambda_{\alpha} = 1215.67\,\ang$.} 
%$\zalp=2.15$ and $\zalp=2.40$, the limits of our desired redshift range.
%Our overall targeting algorithm is as follows:
%First, the field is divided into a uniform grid of $2.75\arcmin$ cells.
%If there is a spectroscopically-confirmed source with $\zbg$ brighter than our nominal survey limit, 
%$\glim$, within the cell, it receives top priority (Priority 1). The next brightest candidate is assigned Priority 3 as a backup.
%If there are no $g<\glim$ spectroscopic sources, the remaining sources are then ranked-ordered by source magnitude
%and then success probability. We assume that faint $g>\glim$ spectroscopically-confirmed sources nevertheless
%have a success probability of $P=1.0$, while photometric redshift candidates are assigned lower probabilities of 
%$P=0.9$ or $P=0.8$ depending on whether they are brighter or fainter than $\glim$.
% The top two sources ranked in this way are assigned Priority 2, while the next two are flagged Priority 3 and Priority 4, 
% respectively.

\begin{figure}\includegraphics[width=0.49\textwidth]{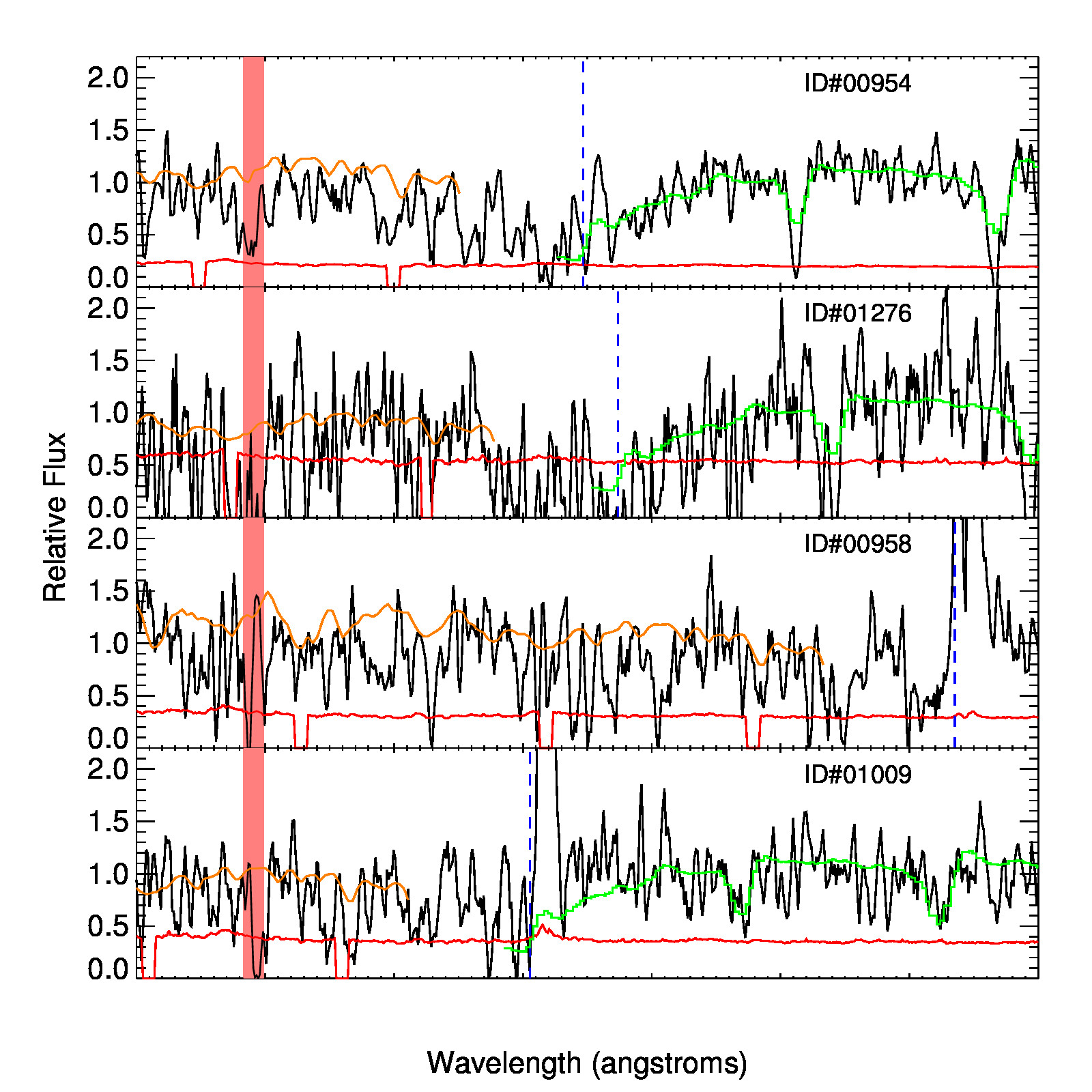}\caption{\label{fig:spec}
Examples of galaxy spectra used for our tomographic reconstruction (smoothed by a 3-pixel boxcar for clarity), 
along with the estimated pixel noise in red, with masked pixels set to zero.
The \lya\ wavelength at
the galaxy redshift is indicated by the vertical dashed-lines, 
while the
green curves shows the LBG composite template from \citet{shapley:2003} for comparison.
All these background sources probe the known $z=2.44$ protocluster, with the
pink shaded region highlighting the \lya\ absorption wavelength ($\lambda\approx4191\,\ang$) at
the protocluster redshift. Strong associated absorption is apparent even in these individual spectra.
Orange curves indicate the estimated continuum level for each object. Note that object \#00958
(3rd panel from top) has a broad \lya\ emission line characteristic of an AGN, but shows 
strong ISM absorption redwards of \lya\ and is thus treated like a galaxy for continuum-fitting.
}\end{figure}

\begin{figure*}\begin{center}\includegraphics[width=0.85\textwidth]{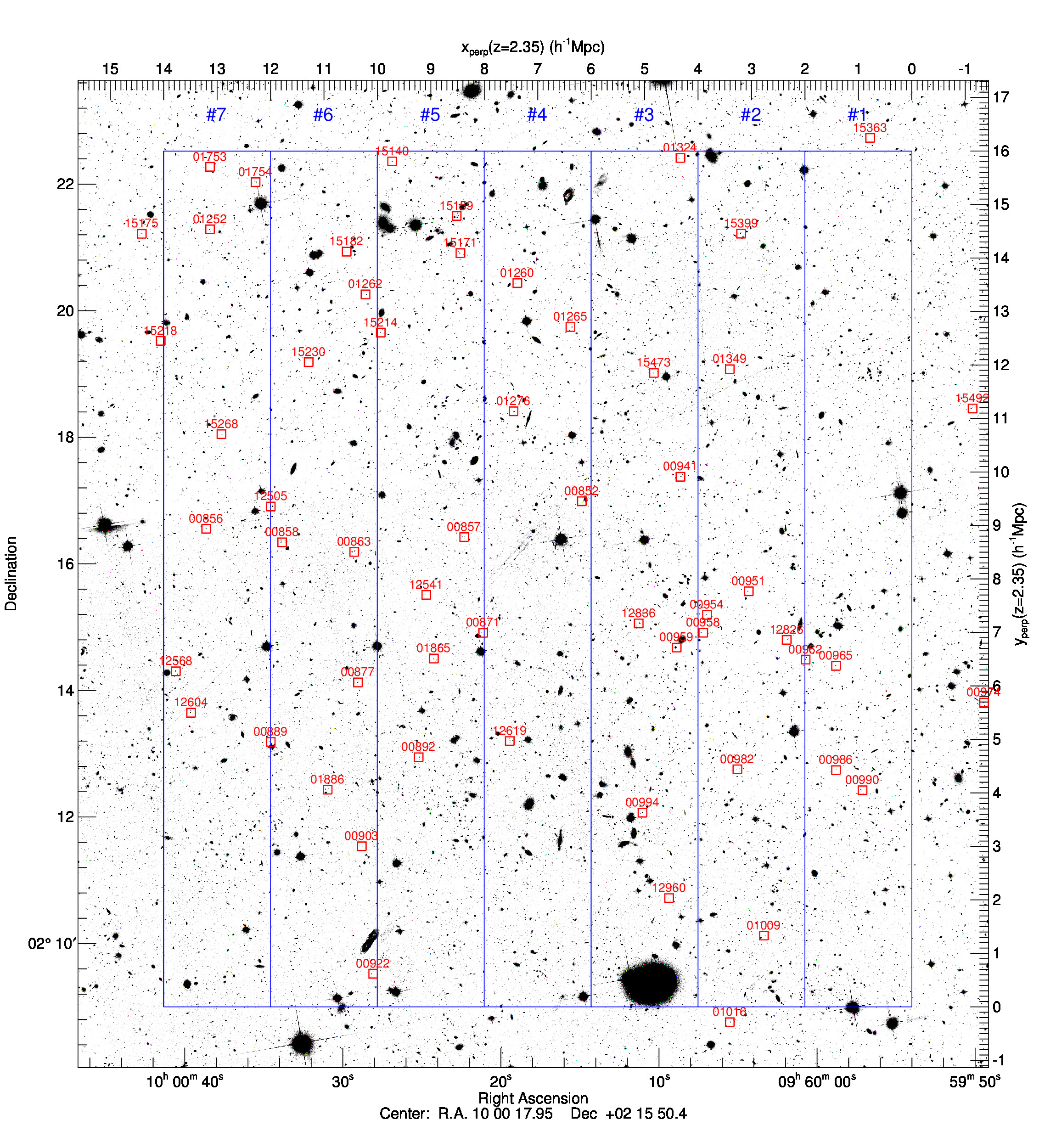}\end{center}\caption{\label{fig:targets}
Hubble ACS image of the COSMOS field \citep{koekemoer:2007} showing the position of our background
galaxies and quasars on the sky, labeled by their ID number. The overall blue box represents the transverse
extent of our tomographic map, while the seven narrow slices indicate the footprint of the map projections shown in 
Figure~\ref{fig:maplong}. The upper- and right-axis labels denote the transverse comoving distance at $z=2.35$, the mean 
redshift of the map.}\end{figure*}

We observed these targets with the LRIS Double-Spectrograph \citep{oke:1995,steidel:2004} on the 
Keck-I telescope at Maunakea, Hawai'i,
%% JFH I think Mauna Kea is two words
during 2014 March 26-27/29-30 and 2015 April 18-20,  
using multi-object slitmasks.
We used the B600/4000 grism on the blue arm and R400/6000 grating on the red with 
the d560 dichroic\footnote{The first night of 2014 observations were taken with the R600/7500 grating
  and d500 dichroic.}, although we work only with the blue spectra in this study. 
With our $1\arcsec$ slits, this yields 
$R\equiv\lambda/\Delta\lambda\approx1000$, which corresponds to a line-of-sight
FWHM of $\approx 3.2\,\hMpc$ at $z\sim 2.3$. 
Despite inclement weather,
%% JFH I would leave out the sub-optimal scheduling issue. Why take a shot
%% at the Keck TAC here?
we managed 
$\sim 15$hrs on-target, observing 7 overlapping masks with 7200-9000s exposure times.
The observations were mostly carried out in $0.5-1\arcsec$ seeing, although one mask was observed in 
$\approx1.4\arcsec$ seeing and consequently had a low yield of useable spectra.
The data was reduced with the LowRedux package in the XIDL  
suite of data reduction software\footnote{\url{http://www.ucolick.org/~xavier/LowRedux/lris_cook.html}}, and
 objects targeted in more than one mask had their spectra co-added.
% this number of 58 objects leaves out one one cpilot12 source which is far away from the map.
We extracted 162 unique spectra, of which 58 objects had secure visual identifications 
 at redshifts $2.3\lesssim z\lesssim3.0$ and therefore
\lyaf\ absorption covering our desired $2.2<z<2.5$ range, as well as adequate S/N ($\snr \geq1.3$ per pixel) 
within the \lyaf. The majority of these sources
are LBGs identified through \lya\ emission or intrinsic absorption, although our sample also includes 4 QSOs, one
of which is our brightest object ($g=20.12$). The faintest object satisfying our S/N criterion was a $g=25.18$
LBG, although this was observed in two masks; the faintest object from a single mask has $g=24.85$.
 
Several example spectra are shown in Figure~\ref{fig:spec}, while the source positions are shown in Figure~\ref{fig:targets}.
The latter also indicates the $11.8\arcmin\times13.5\arcmin$ footprint of our tomographic map. Within the map 
footprint, we have 52 background sources which translates to a projected area density of $\sim1200\,\persqdeg$.
An additional six sources lie just outside the footprint but will be included in the reconstruction input.
The mean transverse separation between the \lya\ forest sightlines probing the foreground IGM is $\dperp \approx 2.4\,\hMpc$. 
%% To do: compute the nlos_eff

We next estimate the intrinsic continua, $C$, of the sources.
For the quasars, we apply PCA-based mean-flux regulation 
\citep[MF-PCA; e.g.,][]{lee:2012a, lee:2013} to the restframe $1041\,\ang<\lambda<1185\,\ang$ 
\lyaf\ region using templates from \citet{paris:2011}, masking intrinsic broad absorption where necessary. 
A similar process is applied on the galaxies, albeit assuming a fixed continuum template from
\citet{berry:2012} and adopting a more generous \lyaf\ range ($1040\,\ang<\lambda<1195\,\ang$).
We also mask $\pm5\,\ang$ around possible intrinsic absorption at \waveion{N}{2}{1084.0}, \waveion{N}{1}{1134.4}, 
and \waveion{C}{3}{1175.7}. L14b estimated an rms error of $\sim 10\%$ for the continua, which is adequate
considering our noisy spectra.

We then compute the \lyaf\ fluctuations, $\delta_F$, 
from the observed spectral flux density, $f$:\beq\delta_F=\frac{f}{C\langle F(z)\rangle}-1,\eeq 
where $\langle F(z)\rangle$ is the mean \lya\ transmission from \citet{faucher-giguere:2008a} evaluated
at redshift $z$. 
The $\delta_F$ values from our background sources, along with the 
associated pixel noise errors $\sigma_N$ estimated by the reduction pipeline, 
constitute a sparse sampling of the IGM \lya\ forest absorption, which we will reconstruct in the next section.

\section{Tomographic Reconstruction}\label{sec:tomo}

\begin{figure*}
\begin{center}\includegraphics[width=\textwidth]{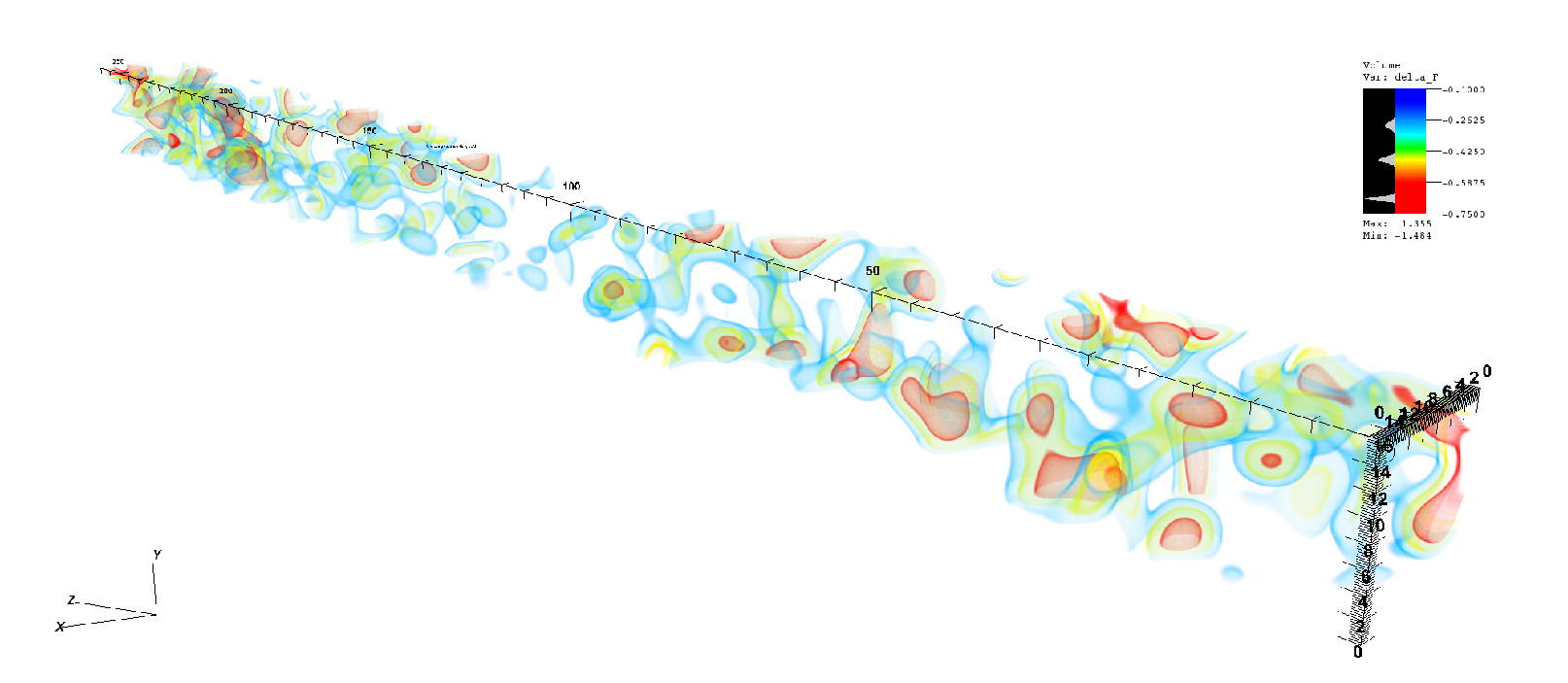}\end{center}
\caption{\label{fig:3d_pano}
Three-dimensional visualization of the tomographic reconstruction from our \lya\ forest data, which can be viewed as a video
online (24MB, \url{https://youtu.be/KeW1UJOPMYI}).
The
$x$- and $y$-dimensions, spanning $14\,\hMpc$ and $16\,\hMpc$ respectively, are on the sky plane (c.f.~Figure~\ref{fig:targets}), 
while the $z$-axis spans $260\,\hMpc$ 
along the line-of-sight. All axes are in units of $\,\hMpc$ comoving, with the $z$-axis originating at $z=2.20$ and extending to $z=2.50$.
Since this map traces \lyaf\ transmitted fraction, more negative values correspond to higher-density
regions. In the video, first 5 seconds show only the sightline positions before the actual map is revealed.}
\end{figure*}

For the tomographic reconstruction, 
we define a 3D output grid with cells of comoving size $0.5\hMpc$, spanning a transverse comoving length 
at $z=2.35$ of $14\,\hMpc$ in the R.A. direction and $16\,\hMpc$ in Dec
(Figure~\ref{fig:targets}).
The output grid spans $2.2<\zalp<2.5$ along the line-of-sight, corresponding to $260\,\hMpc$ with our
simplification that the comoving distance-redshift relationship is evaluated at fixed $z=2.35$.
The overall comoving volume is $V=14\,\hMpc\times16\,\hMpc\times260\,\hMpc=58240\,h^{-3}\mathrm{Mpc}^3\approx(38.8\,\hMpc)^3$, 
i.e.\ nearly $2.5\times$ greater volume than the
L14b map, which corresponds approximately to Slices \#5-7 in Figure~\ref{fig:targets}.

As in L14b, we use a Wiener filtering implementation developed by S15. The reconstructed flux
field on the output grid is given by
\beq \label{eq:wiener}\delrecon=\cmd\cdot (\cdd+\mathbf{N})^{-1}\cdot\delta_F,\eeq
where $\cdd+\mathbf{N}$ and $\cmd$ are the data-data and map-data covariances, respectively.
We assumed a diagonal form for the noise covariance matrix $\mathbf{N}\equiv N_{ii}=\sigma_{N,i}^2$,
and assumed a Gaussian covariance between any two points $r_1$ and $r_2$, such that
$\cdd=\cmd=\mathbf{C(r_1,r_2)}$ and 
\beq\mathbf{C(r_1,r_2)}=\sigma_F^2\exp\left[-\frac{(\Delta r_\parallel)^2}{2L^2_\parallel}\right]\exp\left[-\frac{(\Delta r_\perp)^2}{2L^2_\perp}\right],\eeq
where $\Delta r_\parallel$ and $\Delta r_\perp$ are the distance between 
$\mathbf{r_1}$ and $\mathbf{r_2}$ along, and transverse to the line-of-sight, respectively. 
Again following L14b, we adopt transverse and line-of-sight
correlation lengths of $L_\perp=3.5\,\hMpc$ and $L_\parallel=2.7\,\hMpc$, respectively, 
as well as a normalization of  $\sigma_F=0.8$. These values were found to give reasonable
reconstructions on simulated data sets with similar sightline sampling and S/N as our data.

%It is beyond the scope of this Letter to compute a formal map covariance, but
%we can estimate the relative map sampling at each position $\mathbf{r}$, $\snrvol(\mathbf{r})$, 
%by summing the inverse
%pixel-noise, $1/\sigma_N$, from all spectral pixels within radius $R_{th}$, and comparing with a fiducial 
%sightline configuration:
%\beq\label{eq:mapsig}\snrvol(\mathbf{r})=\int_{V} \frac{\sigma^{\mathrm{fid}}_N (\mathbf{r}')}{\sigma_N (\mathbf{r}')}W_{th}(|\mathbf{r}-\mathbf{r}'|;R_{th})\,\mathrm{d}\mathbf{r}',\eeq
%where $W_{th}$ is the top-hat function and $\sigma^{\mathrm{fid}}_N$ is the pixel noise from an idealized 
%sightline configuration, which we choose to be a uniform grid with $\dperp=3.5\,\hMpc$ separation and
%$\sigma^{\mathrm{fid}}_N=0.4$. Similarly, we choose $R_{th}=3.5\,\hMpc$ to roughly match our
%tomographic reconstruction. Regions with $\snrvol\sim1$ can be considered to be well-sampled.
%% JFH I don't understand why you are mentioning the map error here,
%% or the introducing this S statistic, given that you don't use it
%% until the end of section 4. I would introduce it there where it is used. 

\begin{figure*}\includegraphics[width=\textwidth,clip=true,trim=0 0 0 200]{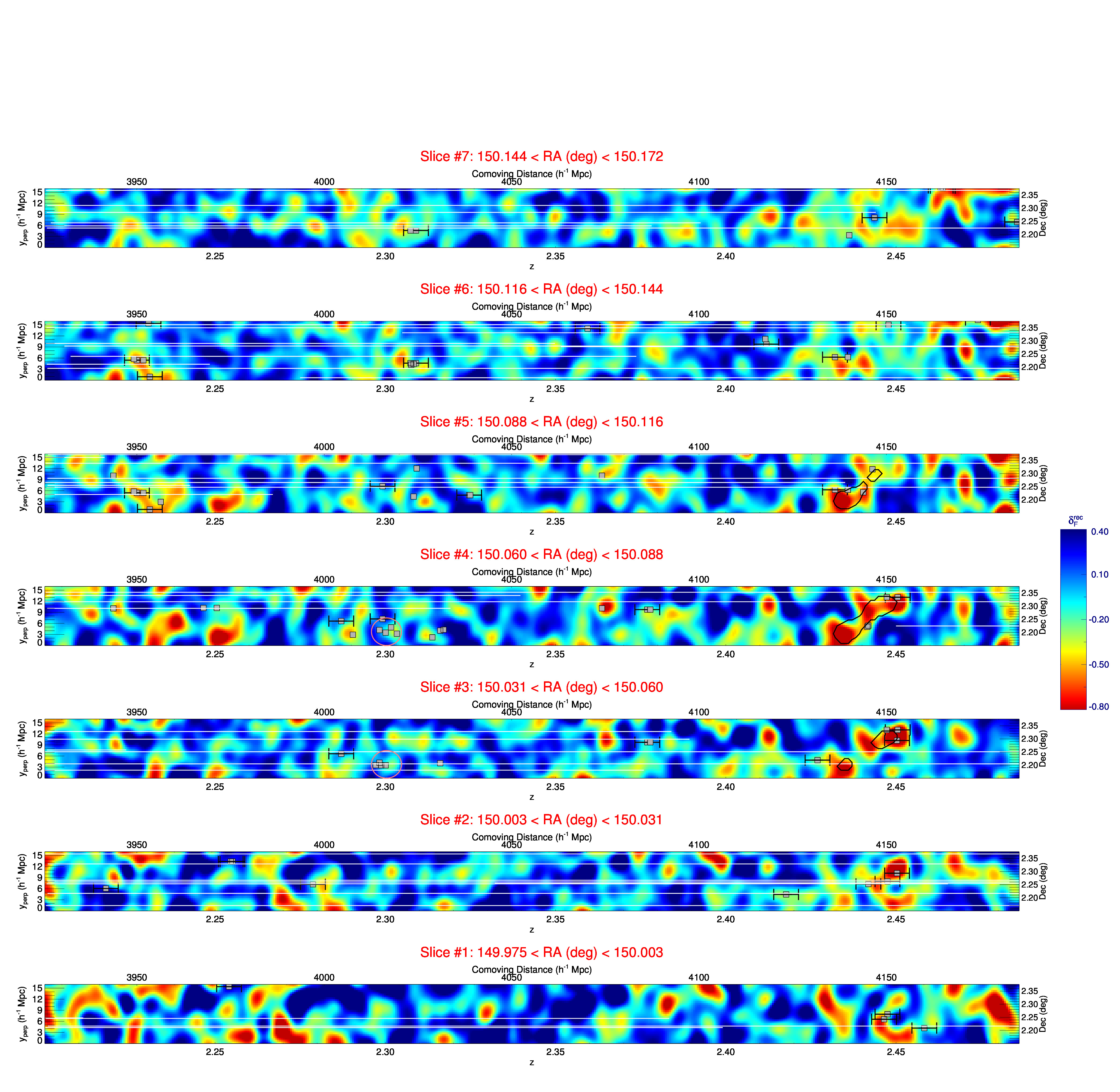}\caption{\label{fig:maplong}
Projections of our \lyaf\ tomographic map across $2\hMpc$ slices in the R.A.\ direction, with the 
slice locations indicated in Figure~\ref{fig:targets}. Note that negative values of $\delrecon$ (red colors)
correspond to higher overdensities, while the white horizontal lines represent the skewer sampling from the
background sources. Grey squares mark the positions of coeval galaxies with spectroscopic redshifts
from the zCOSMOS \citep{lilly:2007} and MOSDEF (K15) surveys. The error bars indicate
the LOS uncertainties associated with LBG redshift determination, while the galaxies without error bars 
have NIR redshifts, in which case the symbol widths denote the LOS uncertainty. 
Black contours in Slices \#3-5 (at $z\approx2.435-2.450$) delineate regions with
$\delsm<-3.5\,\sigsm$, which is the protocluster criterion defined by S15. Pink circles in 
Slices \#3-4 indicate the $z=2.300$ compact overdensity of MOSDEF galaxies within our volume.}\end{figure*}

\begin{figure}\includegraphics[width=0.49\textwidth]{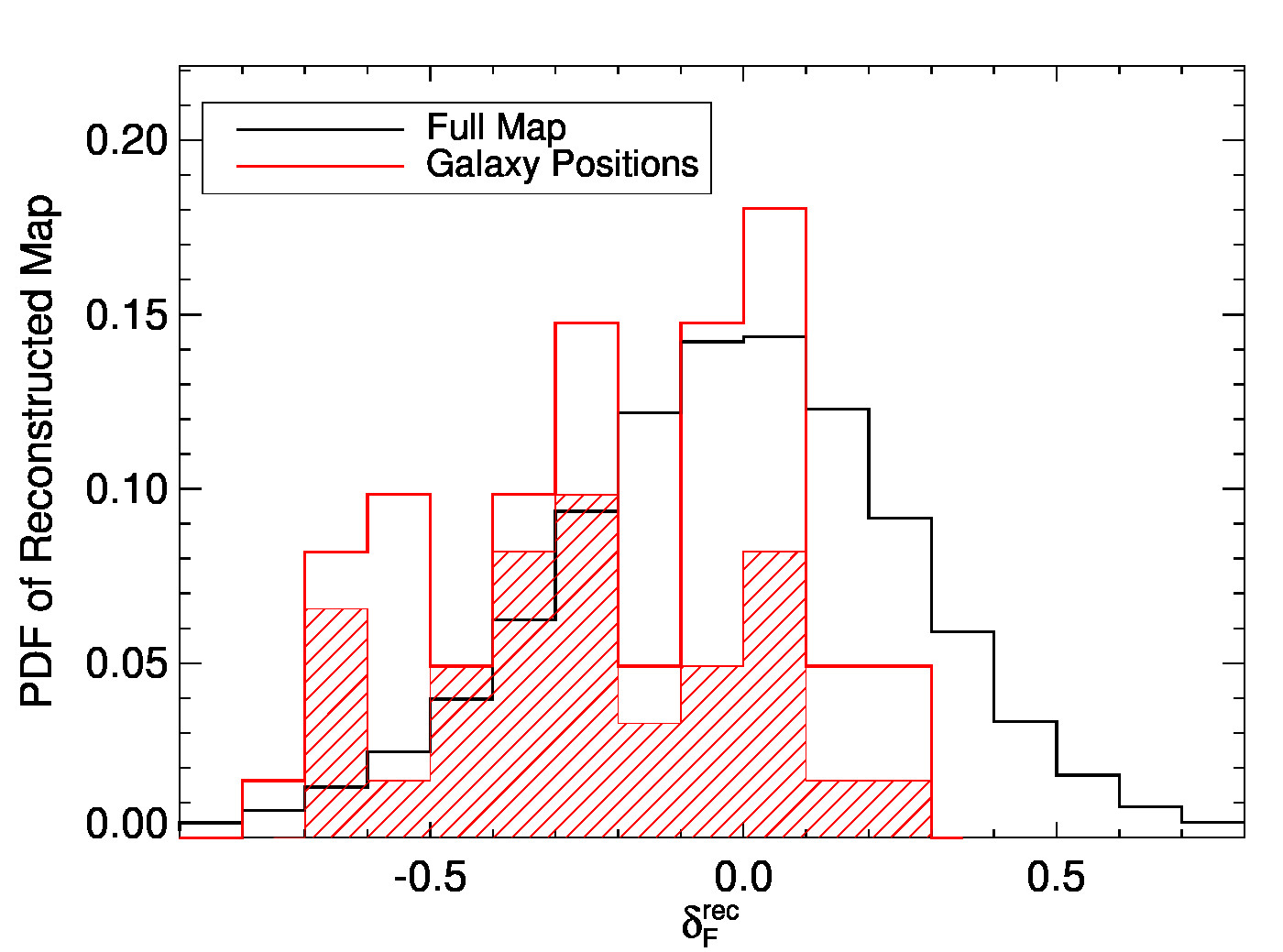}
\caption{\label{fig:maphist}
PDF of the reconstructed absorption from our map (black), along with the corresponding PDF
 at the position of the 61 coeval galaxies (red); negative
absorption corresponds to higher overdensities. The galaxies PDF is clearly skewed towards overdense regions of the
map; this is even clearer in the case of the subsample of 31 MOSDEF redshifts (shaded red histogram).}\end{figure}

The resulting tomographic reconstruction of the foreground IGM is presented in Figure~\ref{fig:3d_pano} as a 3D visualization, 
while Figure~\ref{fig:maplong} shows the same map
as a series of 2D projections across $\Delta\chi=2\,\hMpc$ along the R.A. direction; a movie of the 3D visualization
can be viewed online\footnote{\url{https://youtu.be/KeW1UJOPMYI}}.
%% JFH Should we combine Figure 3 and Figure 12 into the same figure
As in L14b, we see large overdensities and underdensities spanning $\gtrsim10\,\hMpc$. 
The enlarged map volume also allows us to repeat the comparison of the map with coeval galaxies
within the same volume. Using an internal compilation of spectroscopic redshifts within the COSMOS field
(albeit updated from the one used in L14b), we find 61 coeval galaxies,  
primarily from the zCOSMOS-Deep \citep{lilly:2007}, VUDS \citep{le-fevre:2015} and MOSDEF (K15) surveys.
%% JFH Did our spectroscopy uncover any new galaxy redshifts? If so,
%% state that here, i.e. internal compilation from the COSMOS field
%% plus XX from our spectroscopy
%% KG I don't know the best practices for estimating systemic redshifts, and I would also 
%% need to re-inspect everything. So not using those for this paper.
We overplot these galaxy positions on Figure~\ref{fig:maplong}. 
These aforementioned surveys are known 
to be uniformly flux-limited;
%% JFH I guess you want to say ``flux-limited'' or apparent magnnitude limited
we momentarily leave out galaxies targeted specifically as protocluster members,  
which will be discussed in Section~\ref{sec:proto}.

We compare the distribution of map values at the galaxy positions with the overall map PDF in Figure~\ref{fig:maphist}. 
There is a clear skew in the absorption at the galaxies' positions towards negative $\delrecon$ values, confirming that
both the tomographic map and galaxy positions sample high-density regions. 
A two-sample Kolgomorov-Smirnov test finds $P=5.7\times10^{-5}$ that the galaxy values are drawn from
the overall distribution; this is a considerable improvement from the analogous comparison with 18 coeval galaxies in L14b, 
which had $P=0.24$.

There are several effects, however, that can introduce scatter into this comparison between galaxy positions and the
large-scale IGM. Even in the absence of map reconstruction errors and with perfect redshift determinations, one expects
 a slight mismatch from the slightly different velocity fields traced by galaxies and the \lya\ forest, which is in turn due to the
 different halo biases of both tracers.
%% JFH I think the right way to say this is the the IGM and the galaxies trace different velocity fields, i.e. have
%% a different velocity bias, which may be expected given that the galaxies are a more highly biased tracer of the
%% density field than the IGM (ask Martin about references here). 
However, in our case we expect the primary sources of scatter to be reconstruction errors in the tomographic map, as well
as line-of-sight positional uncertainties ($\sigma_v\sim300\,\kms$ or $\sigma_{\mathrm{los}}\sim3.5\,\hMpc$) 
on the galaxies comparable to our map resolution. The latter arises from redshift errors induced by
%% JFH I added in the word redshift error
low spectral resolution ($R<200$)  
\citep{lilly:2007,le-fevre:2013} as well as intrinsic scatter between the true systemic redshift compared with
redshift estimates from \lya\ emission or UV absorption lines \citep[e.g.][]{adelberger:2005,steidel:2010,rakic:2011}.
%% JFH There are two relevant references here that you need to add. The first is
%% by Adelberger et al. 200?. The second is by Steidel et al. 2010 (the Ly-a stacking paper, where you can also find the Adelberger et al. reference). In these papers
%% they discuss the various different redshifts you can get from LBGs. I would also include Rakic who had a nice way of quantifying this, but you should cite these
%% other papers, particuarly Steidel quantifies errors using near-IR redshifts. 
L14b showed using simulated reconstructions that this redshift error, along with reconstruction
noise in the tomographic maps, would scatter some
galaxies from overdensities into adjacent underdensities in the tomographic map.

However, galaxy redshifts measured in the NIR have much smaller
redshift errors \citep[$\sigma_v\sim60\,\kms$,][]{steidel:2010}
%% JFH Did Rakic quantify this directly? You also need to cite the Steidel et al.
%% paper I mentioned, and also one of the MOSDEF papers here.
%% JFH I would write this `` due both to higher resolution and the fact that nebular lines in the rest-frame optical
%% are much better tracers of the systemic frame.''
due both to the higher resolution \citep[$R\sim3500$,][]{kriek:2015} and the fact that rest-frame optical 
nebular lines are much better tracers of the systemic frame, 
so when we restrict the PDF comparison 
only to the 31 galaxies observed in the NIR by MOSDEF, we find an even stronger skew towards higher 
map overdensities: the MOSDEF galaxies sample a median map value of $\delrecon\approx-0.24$, whereas 
the other galaxies have a median of $\delrecon\approx-0.09$.
%% JFH I don't think we necessarily need to work with median here, since the
%% the distributions are pretty well behaved. If you work with mean, you can
%% easily quote the errors, and thus we can tell whether these differences
%% are statistically significant, or if they are just due to sampling errors.
%% You can also do that with median, i.e. quote error, but then you need to resort
%% to some kind of bootstrapping. I would suggest you just quote mean values
%% as well as the error. 
% KG I didn't quote the mean because they are both actually pretty similar. 
% The non-MOSDEF galaxies are mostly in lower-density regions but have a small
% number in the very high-density tail which skews their mean.
This comparison is encouraging and indicates that with expanded map volumes, 
 we will be able to study correspondingly larger samples of coeval galaxies as a function of large-scale environment,
 with cuts in various galaxy properties such as star-formation rates, metallicities, AGN activity etc.
 %% JFH Can we also do the KS test for the MOSDEF galaxies, versus the KS test for
 %% all, and note the improvement. Or KS for MOSDEF, KS for not-MOSDEF, and KS for all. 
 
 It is worth briefly mentioning the large regions in our tomographic map with positive \delrecon\ 
 (blue regions in Figure~\ref{fig:maplong}) that are completely empty of galaxies. 
 These likely correspond to the high-redshift voids recently discussed in \citet{stark:2015a}, 
 but we defer a more detailed investigation to a subsequent paper.

\section{Simulations}\label{sec:sims}
Before we proceed to search for overdensities in the COSMOS tomographic map, we first describe the N-body simulations
that we will use to facillitate our subsequent analysis, which are the same simulations as S15.
These are $2560^3$-particle TreePM \citep{white:2002}
simulations with $8.6 \times 10^7 \, h^{-1} M_\odot$ equal-mass particles within a $(256\,\hMpc)^3$ comoving volume.
The assumed cosmology was consistent with \citet{planck-collaboration:2013}: 
$\Omega_{\rm m} \approx 0.31$, $\Omega_{\rm b} h^2 \approx 0.022$,
$h = 0.6777$, $n_s = 0.9611$, and $\sigma_8 = 0.83$. 

Using a friends-of-friends algorithm to identify halos, S15 found 425 halos at $z=0$ with $M\ge 10^{14}\,\hMpc$
that were defined as galaxy clusters.
The progenitor halos for each cluster were traced backwards to $z=2.5$, and their 
center-of-mass at that epoch defined as the corresponding $z=2.5$ protocluster positions.

The velocities and particles at $z=2.5$ were used to generate \lya\ forest spectra with the fluctuating 
Gunn-Peterson approximation \citep{croft:1998,rorai:2013}. From these skewers, S15 also
generated tomographic maps from realistic
mock data sets with various sightline sampling and noise properties.
These reconstructions used the same Wiener-filtering code described in Section~\ref{sec:tomo}.
In this paper, we will work primarily with their simulated reconstructions with average sightline separation 
$\dperp=2.5\,\hMpc$ binned on to a $(1\,\hMpc)^3$ grid, which were designed specifically to match the L14b data set, including the
spectral resolution and \snr\ distribution in the mock spectra.
%% JFH Also mention that they were smoothed to your resolution

Since progenitors of massive $z\sim0$ clusters occupy overdensities on $\sim 5-10\,\hMpc$ scales at
$z>2$ \citep{chiang:2013}, S15 argued that an efficient way to identify $z\sim 2.5$ protoclusters in \lyaf\ tomographic maps is to
first smooth the map with a $\sigma=4\,\hMpc$ Gaussian kernel to obtain the smoothed flux \delsm, and then 
identify regions with $\delsm < -3.5\,\sigsm$,
where  $\sigsm^2$ is the map variance in the respective map after smoothing. We find $\sigsm^2 = \scien{3.5}{-3}$ in
our real map, and $\sigsm^2 = \scien{3.8}{-3}$ the simulated maps.
%% JFH Are you using your map variance for this in the real data or this number from Casey? 

\begin{figure}\includegraphics[width=0.49\textwidth]{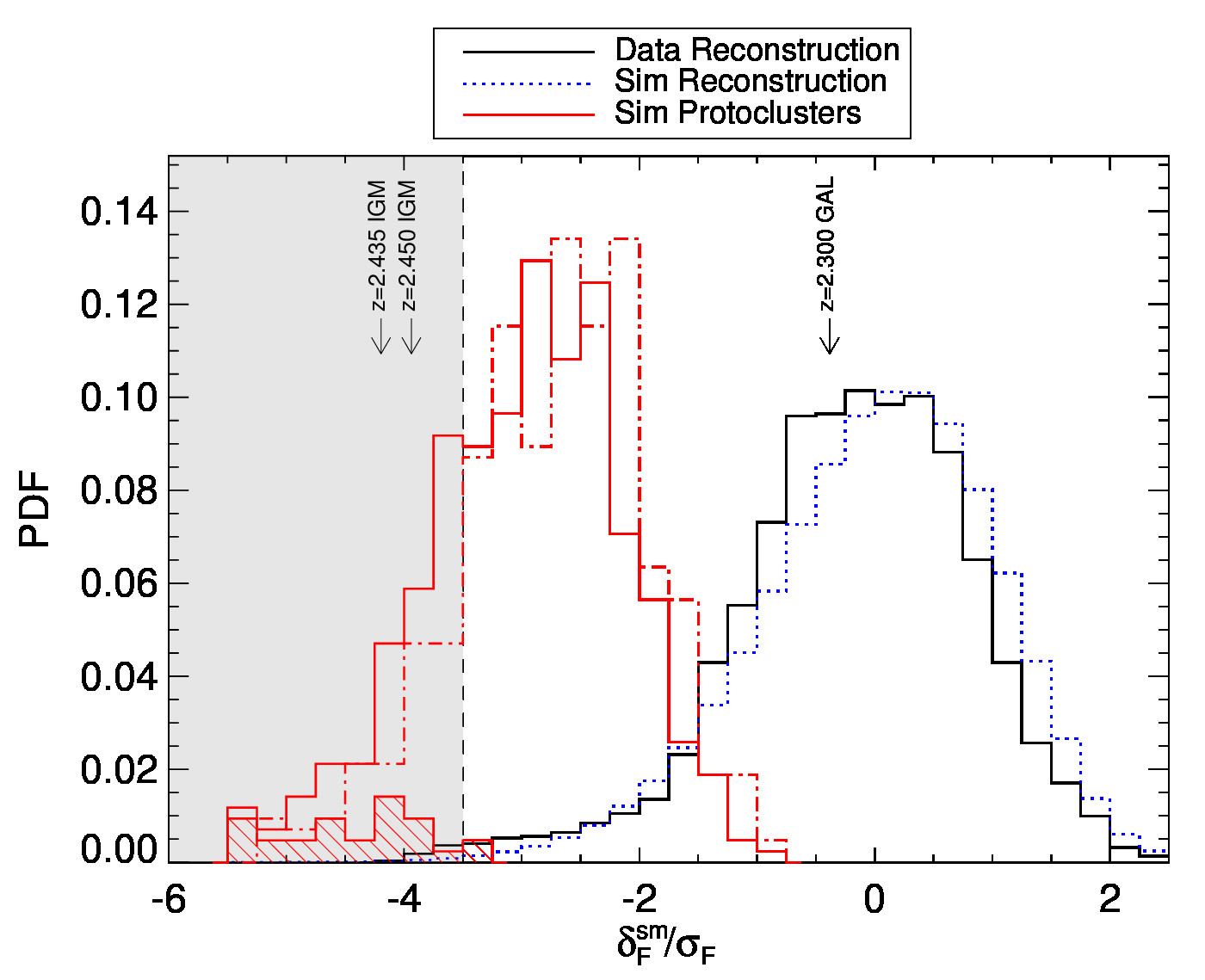}
\caption{\label{fig:threshist}
PDF of the quantity $\delsm/\sigsm$ used to select for protoclusters. 
The black histogram shows the distribution from our real map, while
the blue-dotted histogram shows that from simulated reconstructions.
The distribution corresponding to the 425 $M(z=0)> 10^{14}\,M_\odot$ protoclusters within the simulation is shown by
the red histogram, while the hashed red histogram shows the subset of 27 protoclusters with $M(z=0)> 5\times10^{14}\,M_\odot$.
The dotted-dashed red histogram shows the protocluster distribution including our toy feedback model (see Section~\ref{sec:galoverden}).
The shaded grey region indicates the $\delsm/\sigsm<-3.5$ criterion for selecting protoclusters. Downward arrows indicate
the values corresponding to the $z=2.435$ and $z=2.450$ lobes of protocluster candidate found in our tomographic map, 
as well as the $z=2.300$ overdensity of MOSDEF galaxies.
}\end{figure}

\begin{figure*}
\hspace{1.23in}\includegraphics[height=0.03\textheight]{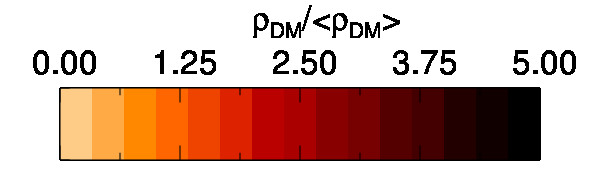}
\hspace{1.64in}\includegraphics[height=0.03\textheight]{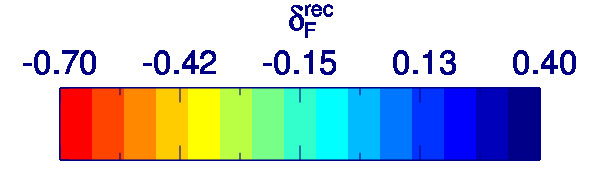}\\  
\vspace{-0.2in}\begin{center}
\begin{overpic}[width=0.8\textwidth]{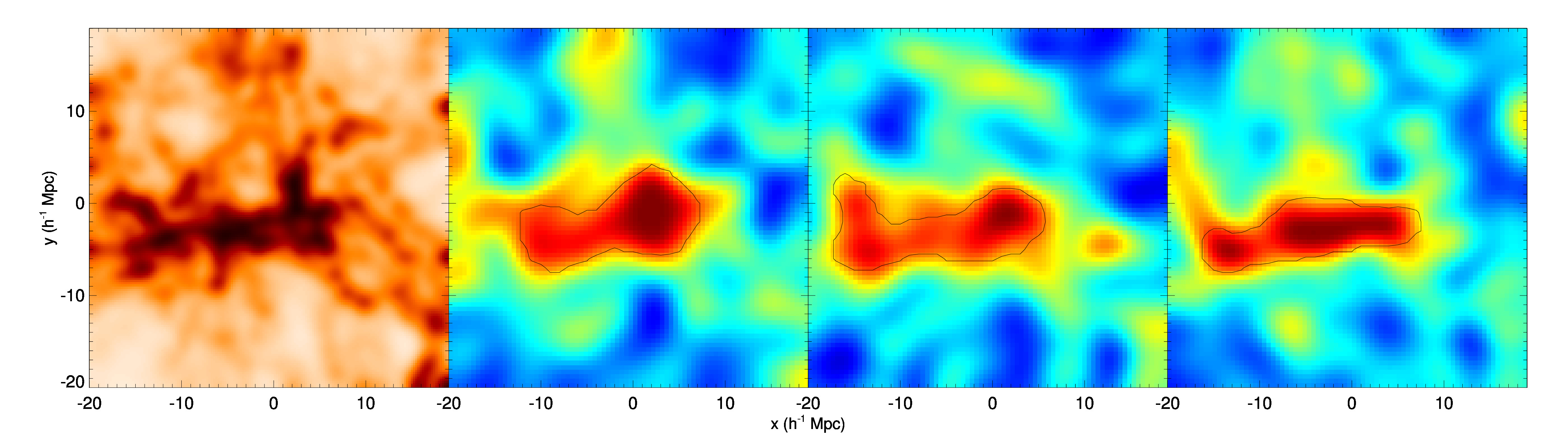}
\put(-6,12){\rotatebox{90}{\color{red} \scriptsize $M(z=0)=10^{15.0}\,\hmsol$}}
\end{overpic}
\\ 
\begin{overpic}[width=0.8\textwidth]{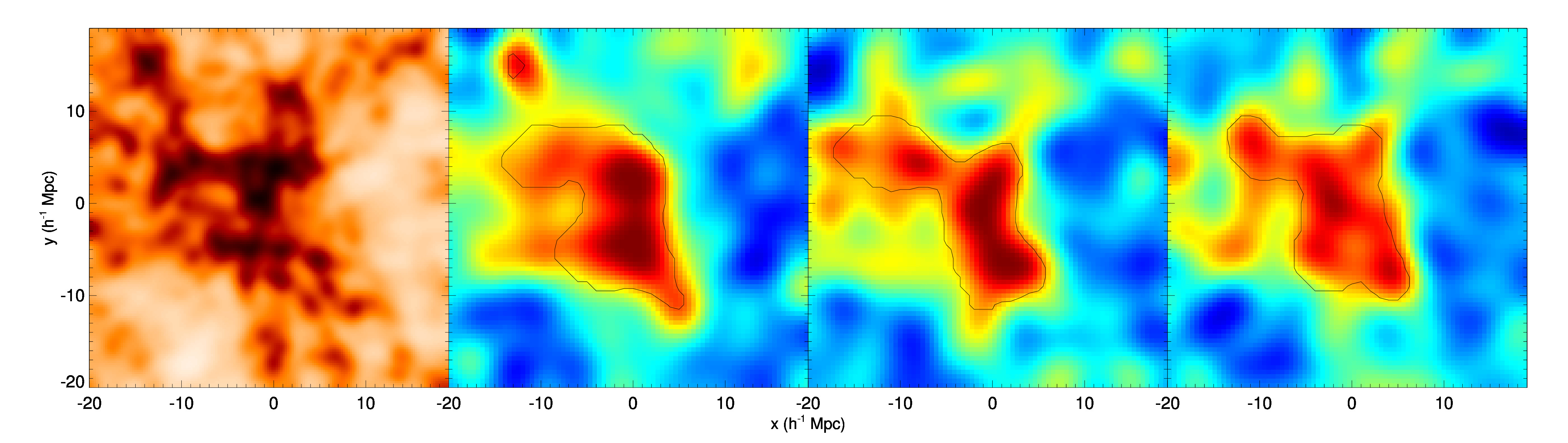} 
\put(-6,12){\rotatebox{90}{\color{red} \scriptsize $M(z=0)=10^{14.9}\,\hmsol$}}
\end{overpic} 
\\
\begin{overpic}[width=0.8\textwidth]{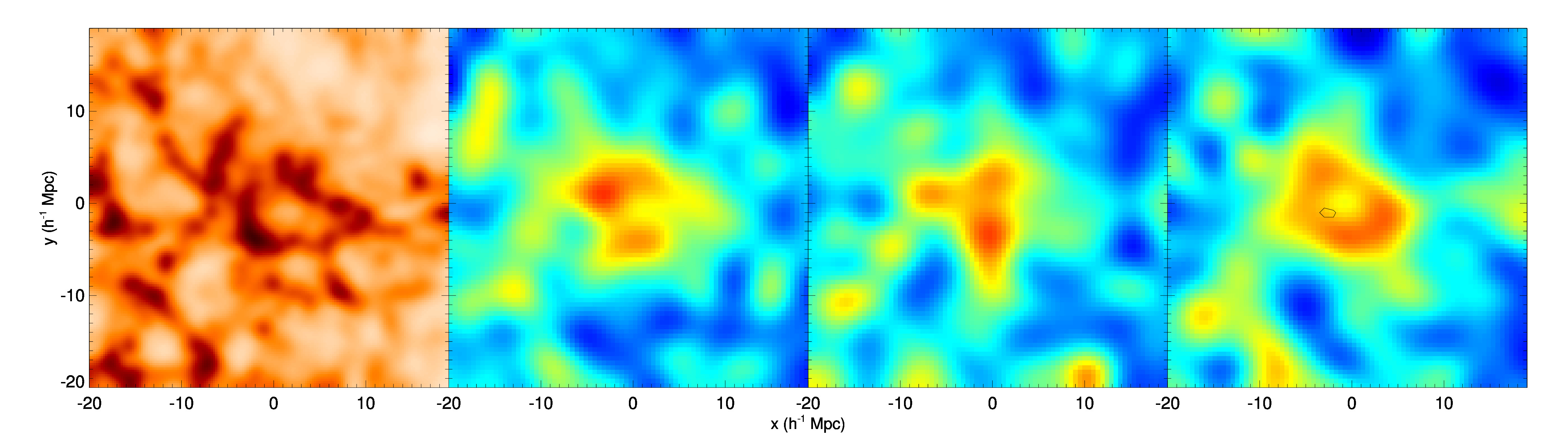}
\put(-6,12){\rotatebox{90}{\color{red} \scriptsize $M(z=0)=10^{14.1}\,\hmsol$}}
\end{overpic}
\end{center}
\caption{\label{fig:protoreal}
Simulated $z=2.5$ galaxy protoclusters shown as projections on the sky plane with thickness $5\,\hMpc$
along the line-of-sight. Each map is centered on the center-of-mass of progenitor halos
that will eventually collapse into $z=0$ clusters with (top to bottom) $M = [1.1\times10^{15}, 8.8\times10^{14}, 1.2\times10^{14}]\,\hmsol$.
The left panel on each row shows the underlying DM distribution (smoothed on $1\,\hMpc$ scales), while the other three panels
show tomographic reconstructions from different realizations of mock \lya\ forest data with similar sightline separations
($\dperp = 2.5\,\hMpc$)  and noise as our data.
The black contours overplotted on the tomographic maps indicate regions that satisfy our protocluster criterion of $\delsm<-3.5\,\sigsm$
after smoothing. While this threshold robustly selects massive cluster progenitors, 
it is marginal for lower-mass protoclusters such as the bottom panel.
Note that in the middle panel, the lobe to the upper left ($[\Delta x,\Delta y]\approx [-10,8]\,\hMpc$) will in fact a collapse into a 
separate, albeit lower mass ($M=1.5\times10^{14}\,\hmsol$) $z\sim 0$ cluster than the central object.}
\end{figure*}

In Figure~\ref{fig:threshist} we show the distribution of smoothed $\delsm/\sigsm$ values from the overall  $(256\,\hMpc)^3$ 
simulation volume (dotted-blue histogram), while the red histogram shows the minimum $\delsm/\sigsm$
within $4\,\hMpc$ of each of the 425 protoclusters within the simulation;
 the shaded histogram shows the subset of 
27 protoclusters that will grow into $M>5\times10^{14}\,\hMpc$ clusters by $z=0$. 
This clearly shows that protoclusters occupy extreme $\delsm/\sigsm$ regions that are well-separated from the
overall distribution\footnote{S15 showed a similar comparison in their Figure~3, but with the `true' \lya\
  absorption field from the simulation, whereas we use the simulated tomographic reconstruction
%% JFH add here ``with noise and resolution matched to the real data''
  with noise and resolution matched to real data.}.

S15 have shown that this method of selecting protoclusters yields good purity:
applying the same selection criteria on the simulated tomographic map with $\dperp=2.5\,\hMpc$, we found 89 candidate 
protoclusters of which 81 will eventually collapse into a $M\geq 10^{14}\,h^{-1}\,\hmsol$ halo by $z=0$ 
($\sim 90\%$ purity);
another 7 candidates ($\sim 8\%$) have group-sized descendants of $3\times 10^{13}\,\hmsol< M(z=0) < 10^{14}\,\hmsol$, 
while only 1 candidate ($\sim 1\%$) will evolve into a low-mass halo ($M(z=0) <3\times10^{13}\,\hmsol$). 
These results are qualitatively similar to those found by S15 but are not numerically identical
despite being applied to the same simulations; this is likely because we have slightly different linking criteria
for identifying contiguous thresholded regions.

%% JFH Why is there a weird grid pattern suprosed on Figure 7 tomo maps??
Figure~\ref{fig:protoreal} shows several of the simulated protoclusters,
in each case comparing three tomographic reconstructions matched to the L14b data,
 but with different random realizations of the sightline positions and pixel noise.
 The progenitors of massive clusters occupy such extended overdensities that they can be robustly detected 
 with our sightline sampling, although at lower masses ($M(z=0)\sim 10^{14}\,\hmsol$) the absorption signature 
 can be less pronounced and is less likely to be selected by our criterion (e.g.\ bottom panel of Figure~\ref{fig:protoreal}).
 
The overall completeness of the S15 selection criterion is $\sim 20\%$ for all $M(z=0)>10^{14}$ cluster progenitors (81/425), 
but this improves dramatically for higher-mass progenitors: we detect 41/51
protoclusters with $M(z=0) > 3\times10^{14}\,\hmsol$ ($\sim 80\%$ completeness) and
24/27 protoclusters with $M(z=0) > 5\times10^{14}\,\hmsol$ ($\sim 90\%$ completeness).
%% JFH Not sure why you are constantly changing threshold, i.e. before you quoted 1 and 3e14 for purity, now you
%% quote 5e14 for completeness. I think it is fine to quote the 5e14 numbers to emphasize how complete you are
%% at high masses, but it is also important to use a consistent set of thresholds. Maybe just add 3e14 here. 
The comoving space density of $M(z=0)>10^{14}\,\hmsol$ clusters is $n \approx 2.5\times 10^{-5}\,h^{3}\,\mathrm{Mpc}^{-3}$
 within our simulation, which translates to an expectation of $\sim 1.5$ protoclusters within our 
 present observed map volume ($V\approx \scien{5.8}{4}\,\hMpc$).
 %% JFH This same number of 1.5 should be in the intro where you mention this stuff, and then multiply that by the
 %% completeness to get a probability of ~ 10% there. 
 
%% JFH Given that you just talked about completeness in the previous paragraph, this choice of words is awkward. 
 Figure~\ref{fig:threshist} also shows the smoothed $\delsm/\sigsm$ distribution from our COSMOS
tomographic map  (black histogram), with both the $\sigsm$ values derived individually for the data and simulated maps. 
There is some disagreement in the distributions from the data and the simulations, 
but the \lya\ forest absorption PDF is known to be a notoriously difficult quantity to model: our simulation
used a DM-only
%% JFH I might say ``DM only N-body code that models only gravitational clustering, but not the hydrodynamics of the IGM''
N-body code that models only gravitational clustering, but not the hydrodynamics of the IGM \citep{white:2010}, nor
have we attempted to accurately model systematics such as the continuum-errors, spectral noise, optically-thick absorbers 
\citep[e.g.,][]{lee:2015}, temperature-density relationship, or Jeans smoothing \citep{rorai:2013}.
%% JFH Maybe mention that things like the rho-T relation and Jeans smoothing are treated in approximate ways and
%% cite Rorai's paper. 
For the present analysis, it is of some concern that our map PDF is somewhat larger at $\delsm < -3.5\,\sigsm$ than the simulated PDF, 
but this is likely to be due to damped \lya\ absorbers (DLAs) and 
Lyman-limit systems (LLS's) with column densities of 
$\nhi \gtrsim 10^{17}\,\persqcm$ contaminating individual sightlines.
By adding individual fake DLAs to our simulated reconstructions, 
we found that these could cause small regions with limited transverse
extent ($\lesssim 1\,\hMpc$) to 
cross our protocluster threshold, but they cannot cause extended transverse IGM signatures characteristic
of massive protoclusters (Figure~\ref{fig:protoreal}), since this requires strong correlated absorption across multiple
adjacent sightlines.
Over the total $\Delta z\sim 12$ redshift path-length of our \lyaf\ spectra  
we expect only $\sim 2$ DLAs with $\nhi > 10^{20.3}\,\persqcm$ \citep{prochaska:2005},
which makes it very unlikely that chance alignments of DLAs could cause a spurious protocluster candidate.
Careful forward-modeling of such systematics would be desirable to characterize marginal protocluster candidates
in future data sets, 
but as we shall see we have only one protocluster candidate in our data, 
which is highly extended ($\gtrsim 10\,\hMpc$) in both the transverse and line-of-sight dimensions.

S15 showed that various quantities related to the \lyaf\ absorption signature of simulated $z = 2.5$ protoclusters can be related
to their $z=0$ descendant masses, although they showed this mostly as a function of the `true' \lya\ absorption field in
the simulation. We now carry out a similar analysis, but incorporating forward modeling 
by using the 81 protoclusters
identified through the tomographic maps of the same simulated absorption field, which incorporate
realistic random sparse sampling of the intervening sightlines as well as noise and spectral resolution matched to our data.
This can then be directly compared with overdensities identified in 
 our data (Section~\ref{sec:igmoverden}).
 In the following discussion, `protocluster candidates' refer to $\delsm<-3.5\,\sigsm$ regions in the 
 $\sigma=4\,\hMpc$ Gaussian-smoothed tomographic maps. Individual protocluster candidates are defined through a 
 $4\,\hMpc$ linking length regardless of physical continuity, so it is possible for a single candidate to be comprised
 of multiple disconnected structures, so long as their voxels can be linked to within $4\,\hMpc$. 
%% JFH Somewhere in the paragraph above use the term ``forward model''
 
 \begin{figure}
\includegraphics[width=0.49\textwidth]{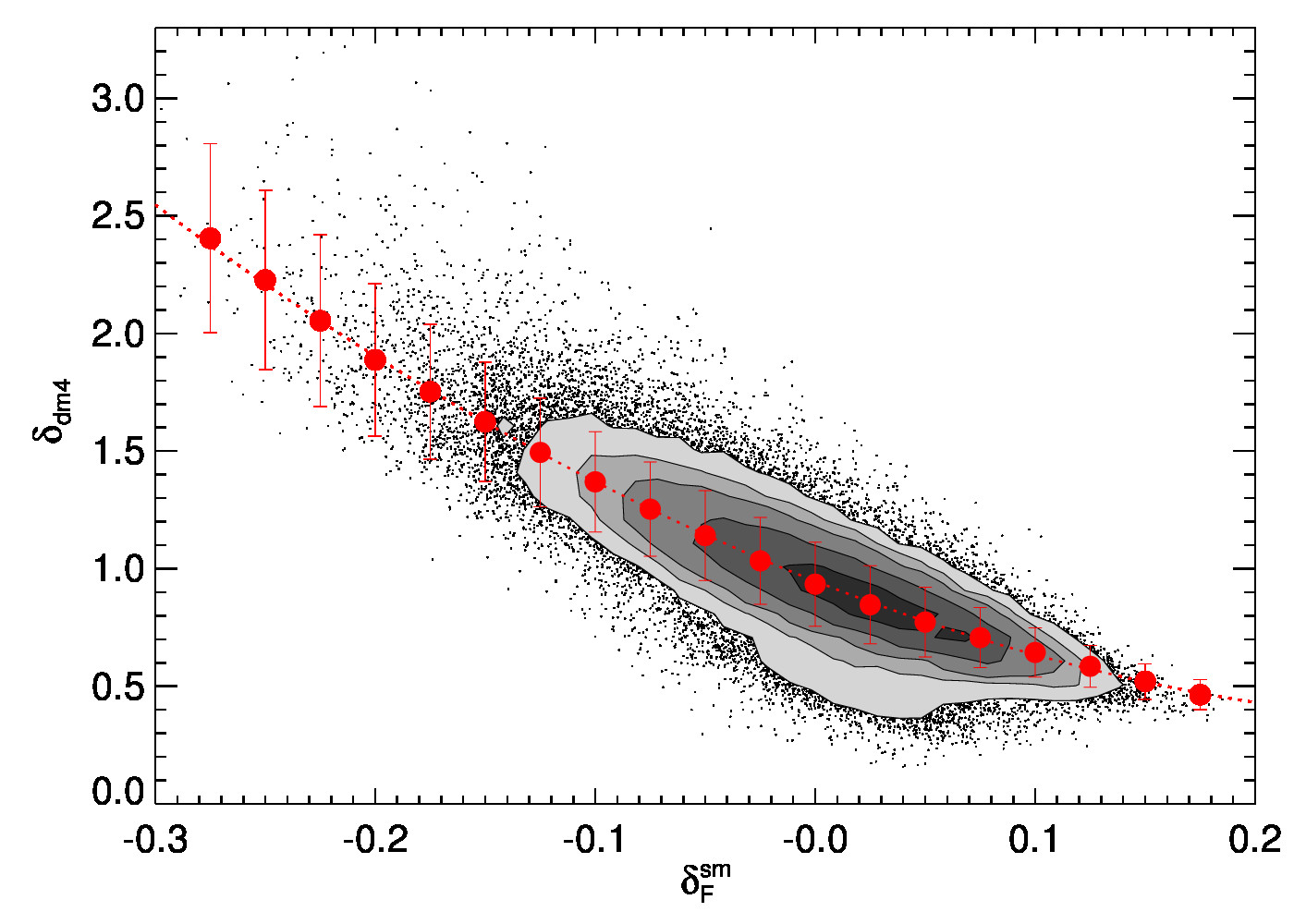}
\caption{\label{fig:fluxmass}
Scatter-plot of the smoothed \lyaf\ tomographic map flux in our simulation against underlying  
DM overdensity smoothed on the same $4\,\hMpc$ scale, evaluated in $(1\,\hMpc)^3$ map voxels. 
The contours denote 
the 10th, 20th, 30th, 50th and 80th percentiles of the distribution, while the red circles denote
the means within bins of $\Delta \delsm = 0.02$ along with the standard deviations as error bars. 
The red-dotted line shows the 2nd-order polynomial fitted to the distribution (Eq.~\ref{eq:fluxmass}). 
}
\end{figure}

\begin{figure}
\includegraphics[width=0.49\textwidth]{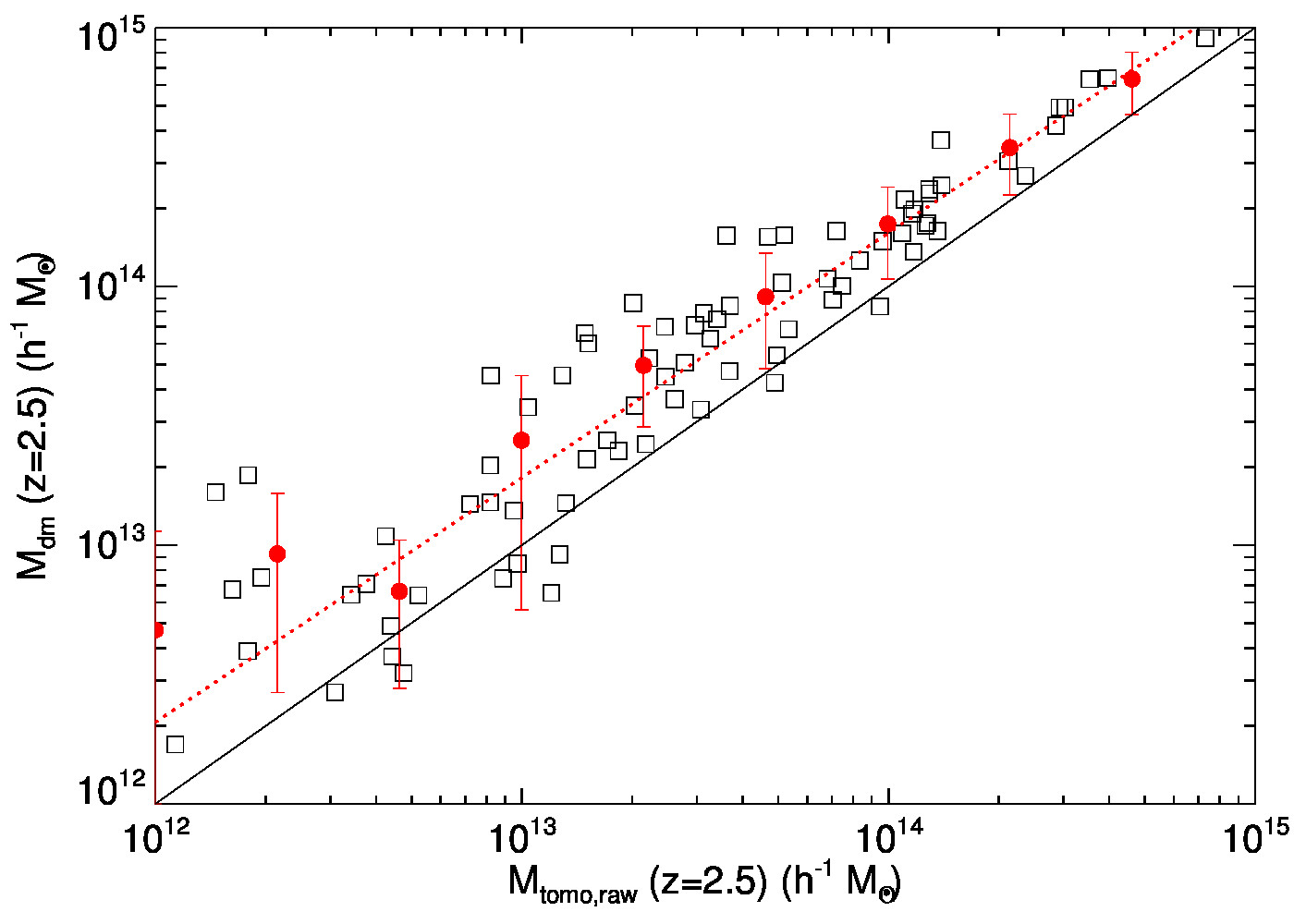}
\caption{\label{fig:mtomo_mdm}
Tomographic mass of protocluster candidates in our simulation, compared with the true DM mass within each
protocluster region at the same redshift of $z=2.5$. The solid black line shows the 1:1 linear slope, 
while the dotted red line shows the best-fit
power-law fit to correct the tomographic masses (Eq.~\ref{eq:mtomo_dm}). 
}
\end{figure}

\begin{figure}
\includegraphics[width=0.49\textwidth]{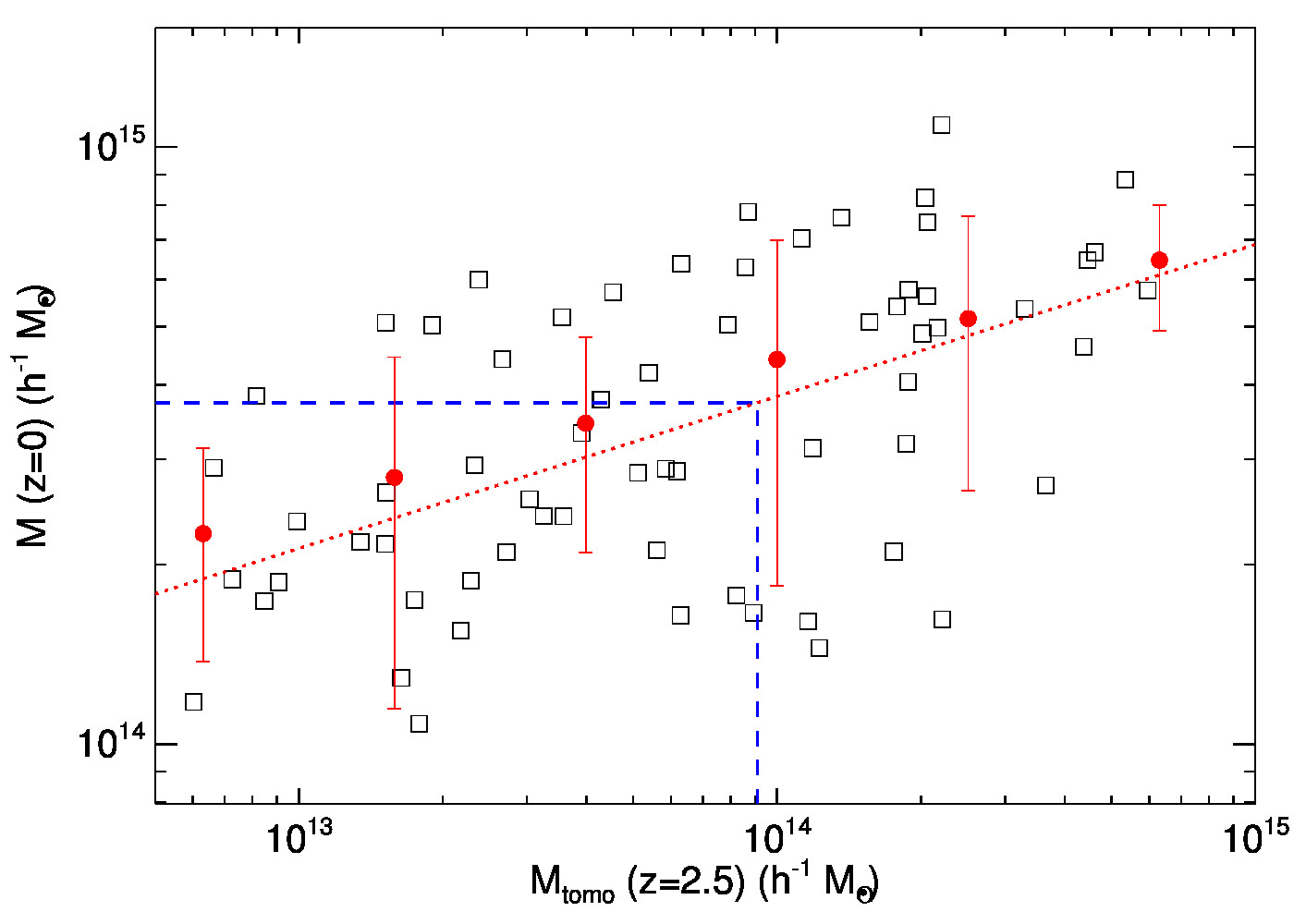}
\caption{\label{fig:mtomo_m0}
Descendant cluster mass at $z=0$, as a function of the corrected tomographic mass (Eq.~\ref{eq:mtomo_dm}) in
$z=2.5$ protoclusters detected in simulated tomographic maps. Red points show the mean and standard deviation in 
logarithmic bins of $0.4\,$dex in $M_{\rm tomo}$, while the dotted red line shows the best-fit power-law
through the points (Eq.~\ref{eq:mtomo_m0}). The blue dashed line indicates the value corresponding to our
observed $z\approx 2.45$ IGM overdensity (Sec.~\ref{sec:igmoverden}).
}
\end{figure}
 
 Since the \lya\ forest absorption is a tracer of the underlying dark matter (DM) density, 
 it should be possible to estimate the total $z\sim 2.5$
mass encompassed by IGM protocluster candidates. While a full inversion of the underlying matter density field from IGM
tomographic maps is a challenging problem beyond the scope of this work 
\citep[although see][]{nusser:1999,pichon:2001,kitaura:2012}, we can use our simulations for the more limited goal of 
estimating the $z=2.5$ mass enclosed within the protocluster regions.
This exploits the tight relationship found by \citet{lee:2014} between the map flux after smoothing by 
$\sigma=4\,\hMpc$ Gaussian, $\delsm$, and
%% JFH Clarify what you mean here by delta_smooth, I got confused. 
$\deldm$, the underlying DM overdensity smoothed to the same scale.
Figure~\ref{fig:fluxmass} shows this relationship from our simulation (binned in $(1\,\hMpc)^3$ voxels), 
which can be fitted with a 2nd-order polynomial:
\beq\label{eq:fluxmass}
\deldm \approx 5.502\,(\delsm)^2 -3.681\,\delsm + 0.947.
\eeq 

We then apply this mapping to the $\delsm$ values within $z=2.5$ tomographic map voxels 
encompassed by the $\delsm/\sigsm<-3.5$ protocluster
%% JFH Make it clear, are you applying this trnasformation to the original map or the 4mpc smoothed tomo map??
contour,
%% JFH By region do you mean contour here? The black line you draw. Make this more clear. 
and then sum these DM overdensities over the protocluster candidate (`pc') region 
to define a raw `tomographic mass':
\beq\label{eq:mtomo}
M_{\rm tomo,raw} = \left(\sum^{\rm pc} \deldm\right) \times \; 8.1\times 10^{10} \hmsol,
\eeq
where $M=8.1\times 10^{10} \hmsol$ is the cosmic mean density of DM and baryons multiplied by the
 $(1\,\hMpc)^3$ comoving volume of our map voxels.
%% JFH Where does the 8.1e10 normalization constant here come from. Please explain that. Should be something like
%% the volume of your cell times the mean density of the Universe. 
Figure~\ref{fig:mtomo_mdm} shows the estimated $z=2.5$ raw tomographic masses for the protocluster candidates
in our simulation against the `true' unsmoothed DM mass within each protocluster region. 
There is a reasonably tight relationship between $M_{\rm dm}$ and $M_{\rm tomo,raw}$, 
but the latter systematically underestimates by $\sim 20-30\%$ the true DM mass over the 
$\sim 10^{13-14}\,\hmsol$ range typical of our protocluster regions, because it was calibrated to the 
smoothed DM field and not the true DM field (binned in $(1\,\hMpc)^3$ cells in our simulation).
Since the protocluster boundaries (defined by $\delsm/\sigsm<-3.5$) are still significantly overdense, 
the smoothing smears DM out of the protocluster regions hence causing an underestimate of protocluster mass.
To get an accurate estimate of the $z=2.5$
DM mass enclosed within any protocluster candidates identified in a tomographic map, we would therefore need to 
correct the raw tomographic masses by the power-law found to fit Figure~\ref{fig:mtomo_mdm}:
\beq\label{eq:mtomo_dm}
M_{\rm tomo} \approx 10^{14.21} \left(\frac{M_{\rm tomo,raw}}{10^{14}\,\hmsol}\right)^{0.946}  \,\hmsol.
\eeq
This is the formula we use when we henceforth refer to `tomographic masses'. Since this calibration is
based on tomographic maps with realistic sightline sampling and noise, the scatter in Figure~\ref{fig:mtomo_mdm}
provides a measure of the uncertainty in the protocluster mass estimate: for example, the
uncertainty in the mass estimate for a $M_{\rm tomo}(z=2.5) = 10^{14}\,\hmsol$ protocluster is 
$\sigma_{\rm tomo} \approx \scien{5}{13}\,\hmsol$. 
%% JFH Here state that ``this is what we do when we define tomographic masses, i.e. we use this formula. Finally
%% quote a number for the scatter of this mass against the true mass, and make some statement like, we see that
%% tomography with our S/N and and resolution provides a method for measuring proto-cluster masses to an accuracy of
%% XX%.

%% JFH As I understand it, this analysis here above was done on mock
%% data, i.e. with S/N and resolution effects, and sparse sampling
%% effect, and not perfect data. Make that clear by explicitly stating
%% that somewhere.

As suggested by S15, we also checked if the $z=2.5$ tomographic protocluster mass could be used to 
predict the $z=0$ descendant mass. This is shown for our simulated protocluster candidates 
in Figure~\ref{fig:mtomo_m0}, where we see a trend for increasing $M(z=0)$ with $M_{\rm tomo}$, which 
can be fitted with the following power-law:
\beq\label{eq:mtomo_m0}
M(z=0) \approx 10^{14.5} \left(\frac{M_{\rm tomo}}{10^{14}\,\hmsol}\right)^{0.25}\,\hmsol.
\eeq 
Part of the $\sim 0.2\,$dex scatter in this relationship is due to the reconstruction
error in our simulated tomographic maps (since we have realistic pixel noise and sightline sampling),
but S15 have also shown that the dominant contributor to this scatter is in fact intrinsic, i.e.\
due to diversity in the morphologies of cluster progenitors at fixed $M(z=0)$.
%% JFH At the end of the day I want to know what the correlation is between M_dm(z =2.5) and M_dm(z=0), what the
%% scatter is in that relation, and I also want to know the correlation between M_tomo(z=2.5) and M_dm(z=0) and the
%% correlation and scatter in that relation. I also want to know whether the dominant source of scatter between
%% ancestors and descencdants is due to tomograhic errors (I doubt it), or intrinsic due to the different
%% paths structure formation can take (I think this is the case). As such, I don't think we need to show the M_dm(z=0) vs
%% M_dm(z=2.5), but it would be nice to quote that scatter, and then empahize that the dominant source of scatter in
%% FIgure 10 is actually structure formation and not tomography. 

%% JFH Rather than M_tomo and M_tomo_corr, I'd just define a M_tomo_raw and M_tomo. Then your axis labels on the
%% plots will be simpler. 

%% JFH This equation here is hanging in the middle of nowhere, with no text referencing it??

\begin{figure}
\includegraphics[width=0.49\textwidth]{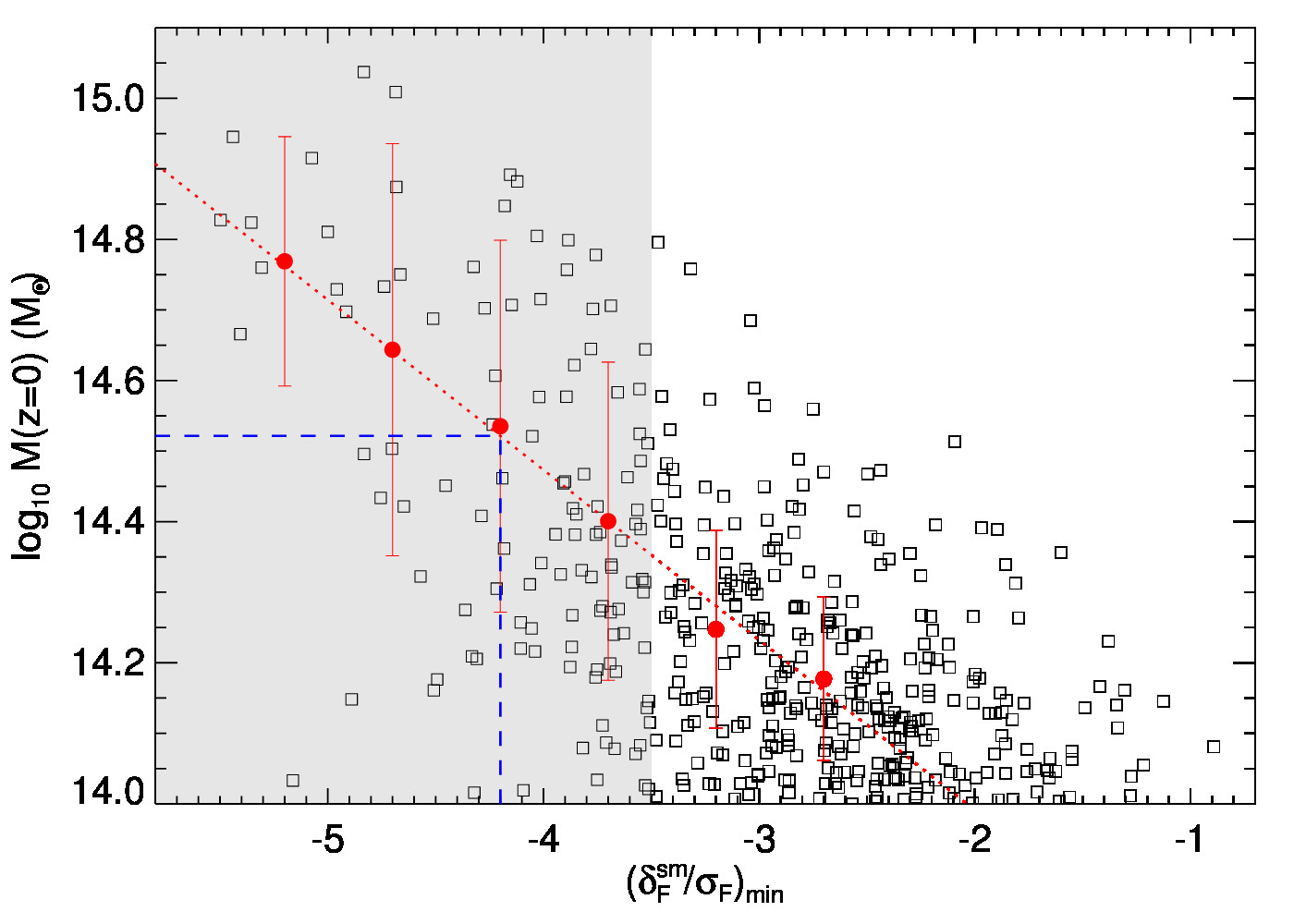}
\caption{\label{fig:delmin_mass}
Relationship between the smoothed absorption peak associated with all $z=2.5$ protoclusters in our simulated tomographic
map, and the final $z=0$ cluster mass.  The red points with error bars show the average $M(z=0)$ in bins of 
$\Delta (\delsm/\sigsm)_{\rm min} = 0.5$
and the associated standard deviations, while the dotted-blue line is the relationship fitted to the distribution (Eq.~\ref{eq:delmin_mass}). 
The shaded gray region is our $\delsm/\sigsm<-3.5$ criterion used for 
protocluster selection in the tomographic maps. Blue dashed lines indicate the value corresponding to the observed
$z\approx 2.45$ IGM overdensity in our map (Sec.~\ref{sec:igmoverden}).
}
\end{figure}

Another quantity shown by S15 to correlate with $z=0$ cluster mass is the IGM absorption peak associated 
with the protocluster candidates in the 
smoothed tomographic maps. Figure~\ref{fig:delmin_mass} shows this for our simulated protocluster candidates.
Again, we see a relationship between the absorption depth $(\delsm/\sigsm)_{\rm min}$ and the final $z=0$ cluster mass, although 
there is more scatter than the analogous plot in S15 (Figure~5 in their paper) since we are working with realistic reconstructed
tomographic maps rather than the `true' absorption field in the simulation. 
We fit the following function to the distribution:
\beq\label{eq:delmin_mass}
 M(z=0) \approx 10^{13.5 - 0.24\,(\delsm/\sigsm)_\mathrm{min}}\,\hmsol.
\eeq
This absorption depth provides an alternative to $M_{\rm tomo}$ as an estimator for $M(z=0)$, although 
both quantities are not completely independent.
%% JFH Maybe comment on which does a better job, which I think is M_tomo by quoting the relative scatter. Or just saying
%% that M_tomo has less scatter. 

\section{Map Overdensities}\label{sec:proto}

In this section, we analyze our reconstructed \lya\ forest map (Section~\ref{sec:tomo}) and look
for IGM overdensities, aided by what we learned from the analysis
%% JFH Rather than ``in context of'' I would say ``aided by what we
%% learned from the analysis of simulations in the previous section''
of simulations in the previous section. We will then also discuss 
two overdensities within our volume identified from external galaxy datasets.

\subsection{IGM Overdensities}\label{sec:igmoverden}
\subsubsection{$z=2.45$ Protocluster}

\begin{figure*}
\begin{center}
\includegraphics[height=0.04\textheight]{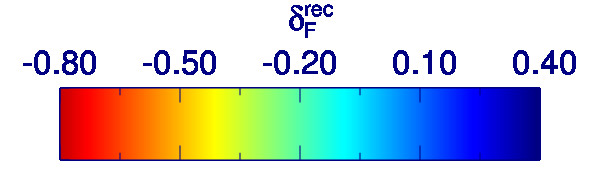}\\  
\begin{overpic}[width=0.43\textwidth,clip=true,trim=0 24 0 0]{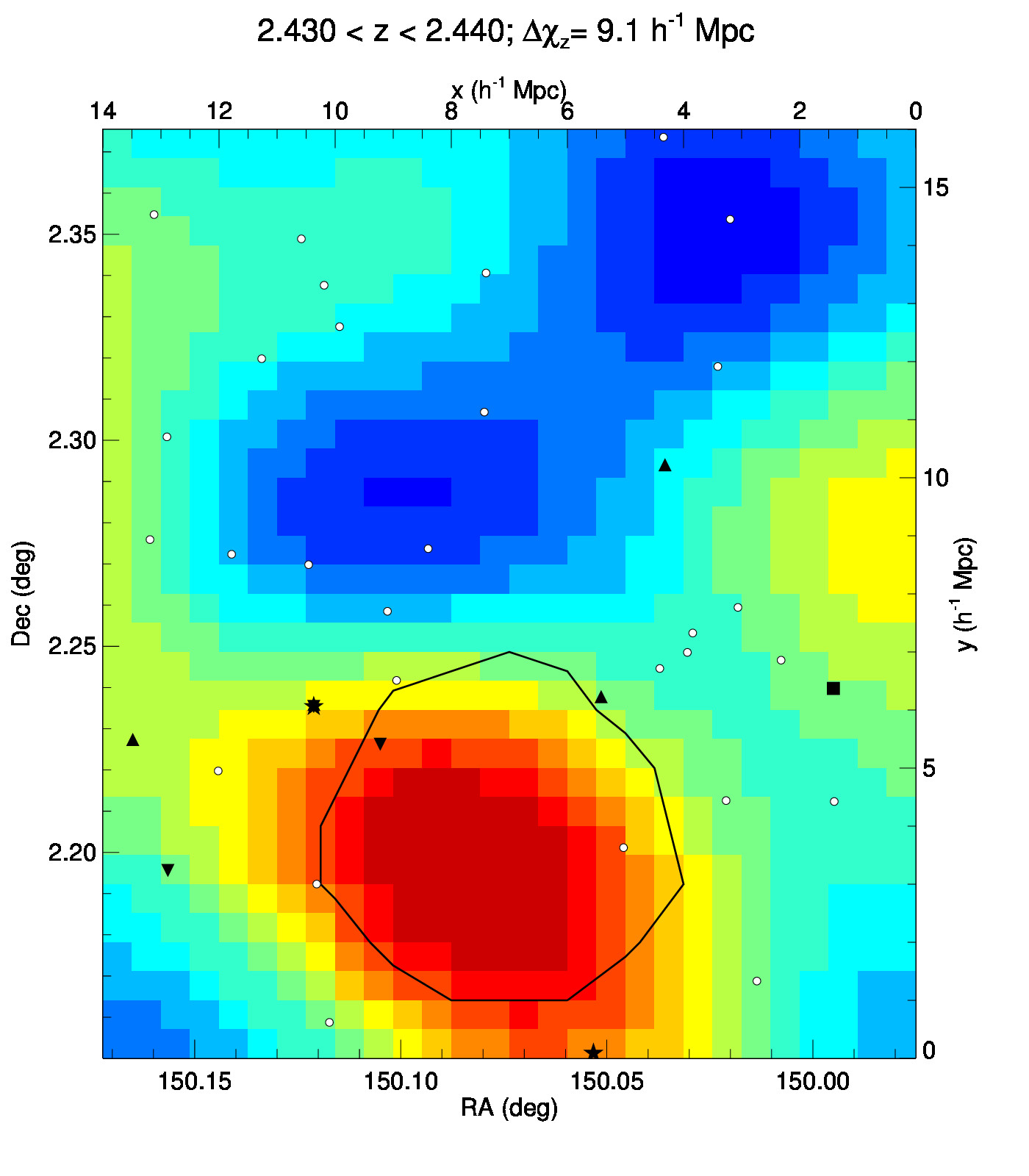} 
\put(42,233){\textbf{(a)}}
\end{overpic}
\begin{overpic}[width=0.43\textwidth,clip=true,trim=0 24 0 0]{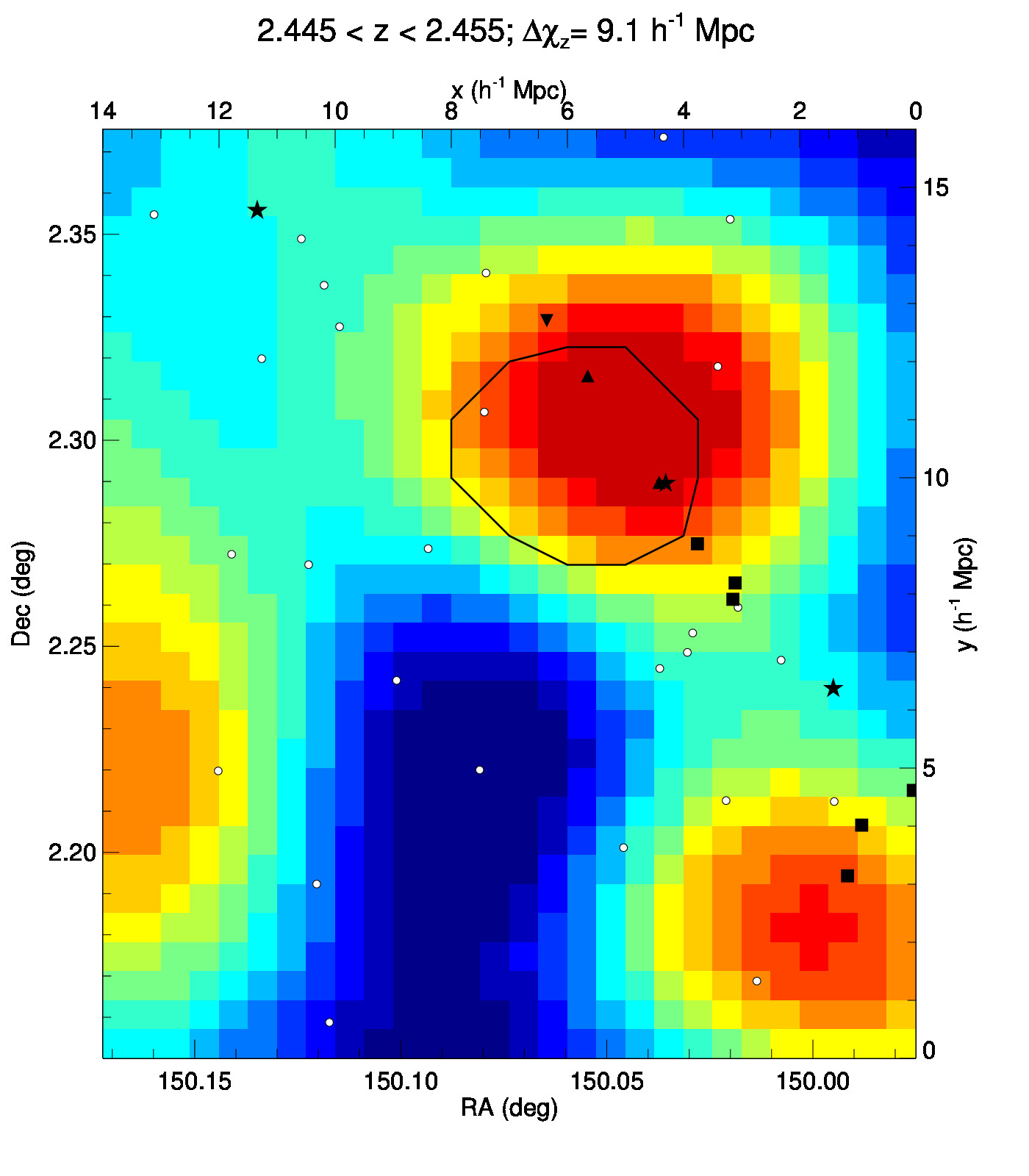}
\put(42,233){\textbf{(b)}}
\end{overpic}
\end{center}
\caption{\label{fig:pcslice1}
Projections of our tomographic map along the line-of-sight, centered on two lobes of the large IGM overdensity
identified through the $\delsm<-3.5\,\sigsm$ protocluster-finding threshold advocated by S15.
The projected redshift range and line-of-sight thickness is labeled above each panel, and were chosen to 
incorporate a reasonable number of coeval galaxies within the projection.
Open circles mark the sightlines sampling each projected slice, while 
filled symbols indicate coeval galaxies within the map: 
MOSDEF galaxies (inverted triangles, K15); zCOSMOS-Deep LBGs \citep[stars,][]{lilly:2007};
LAEs from C15 (triangles); and LBGs from D15 (squares). The black contour denotes the region that satisfies
our protocluster criterion of $\delsm/\sigsm<-3.5$ after smoothing.}\end{figure*}

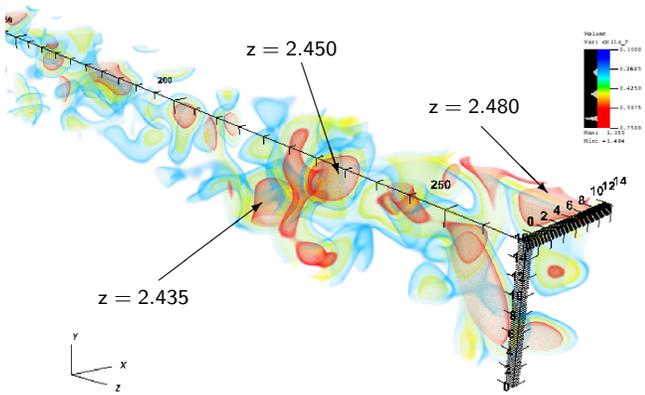
\begin{figure}
\begin{overpic}[width=0.49\textwidth]{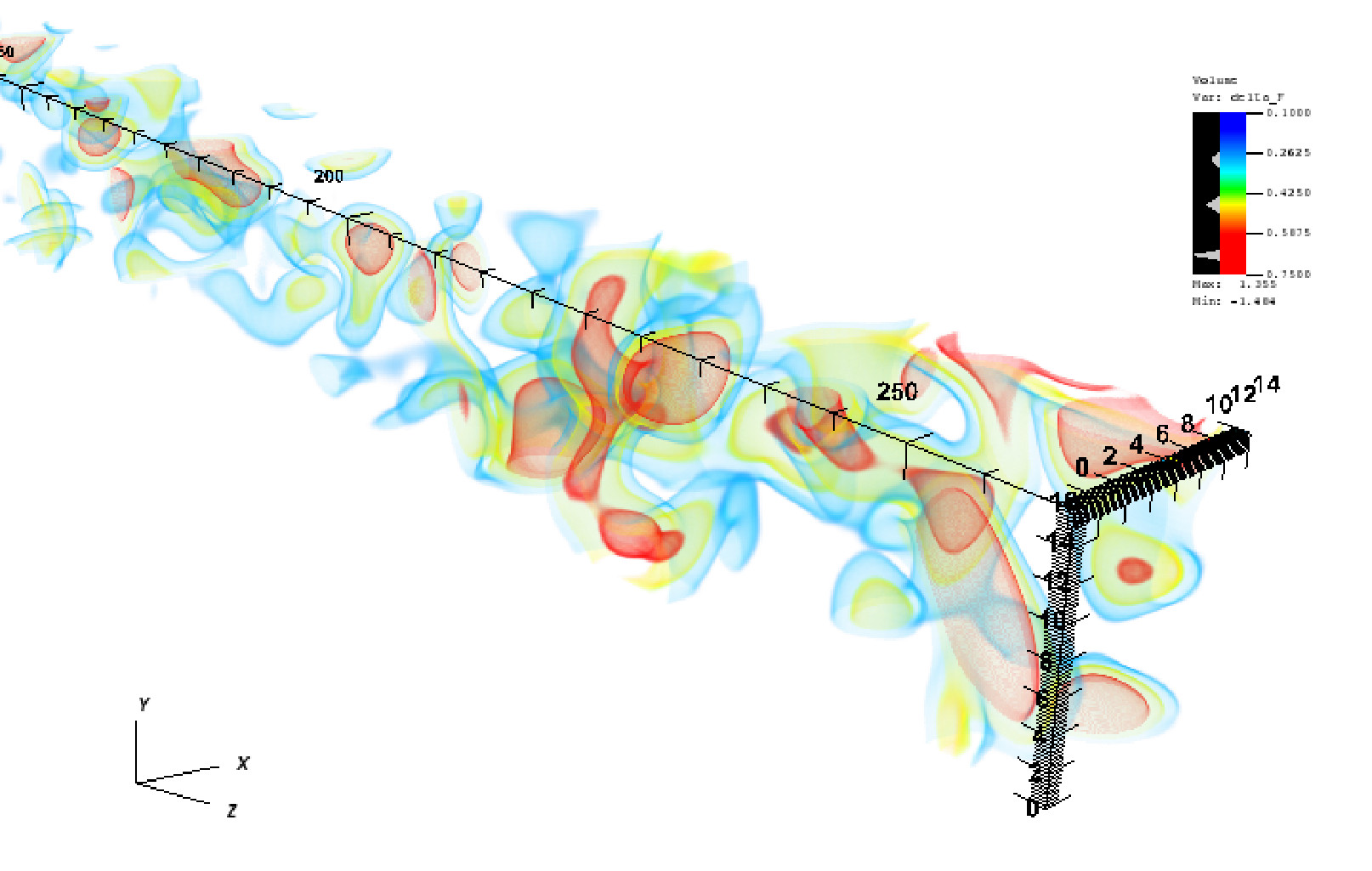}
\put(112,134){\vector(1,-3){14}}
\put(91,137){\footnotesize $\mathsf{z=2.450}$}
\put(180,110){\vector(1,-1){25}}
\put(160,115){\footnotesize $\mathsf{z=2.480}$}
\put(58,52){\vector(4,3){39}}
\put(35,43){\footnotesize $\mathsf{z=2.435}$}
\end{overpic}
\caption{\label{fig:3d_pc}
Still frame from the video in Fig.~\ref{fig:3d_pano} centered on the IGM overdensity at $z\approx 2.44$, with the 
line-of-sight distance ($z$-axis) increasing towards the right. 
Arrows indicate the two absorption peaks ($z=2.435$ and $z=2.450$) associated with the $z\approx 2.45$
protocluster, as well as the $z=2.48$ protocluster from \citet{casey:2015} at the edge of our map (Section~\ref{sec:galoverden}).
}
\end{figure}

We now apply the S15 protocluster criterion on our COSMOS tomographic map.
After smoothing by a $\sigma=4\,\hMpc$ Gaussian kernel, we find one large IGM overdensity within our map
 that clearly satisfies the $\delsm<-3.5\,\sigsm$ protocluster threshold (Figure~\ref{fig:maplong}). The absorption peak 
is at a redshift of $z= 2.450$ and approximately centered on
[RA, Dec]$\approx[150.06, 2.31]\,$deg, although there is a secondary
peak at $z=2.435$ within contour defined by the $\delsm<-3.5\,\sigsm$ threshold; 
these two peaks have smoothed absorption values of $\delsm = [-3.9,-4.2]$ 
at $z=[2.435,2.450]$, respectively, which are also marked in Figure~\ref{fig:threshist}.
%% JFH Given that we have a map, should quote these numbers not as approximate, but with equal signs. 
This overdensity is highly extended in both the line-of-sight and transverse dimensions: 
it can be seen in Figure~\ref{fig:maplong} to continuously extend from $z\approx2.435-2.450$ (Slices \#3-5),
%% JFH add in parenthesis which slice I should be looking at in Figure 4
while
Figure~\ref{fig:pcslice1} shows line-of-sight map projections centered on the two absorption peaks
%% JFH State at the beginning of the paragraph that this is actually two absorption minimna which you link
%% together as one via your procedure. As I just no learn that there are two peaks, and it is a bit confusing. 
of this overdensity
($z= 2.435$ and $z= 2.450$), 
showing that it spans up to $\sim10\,\hMpc$ ($\sim4\,$pMpc) in the transverse direction ---
the overall morphology of the protocluster in the unsmoothed map can be seen in Figure~\ref{fig:3d_pc}.
The total comoving volume of this protocluster candidate (defined as that within the
%% JFH ``defined as that within the del < -3.5 contour''
 $\delsm<-3.5\,\sigsm$ contour) is $V \approx 340\,h^{-3}\,\mathrm{Mpc}^3 \approx (7\,\hMpc)^3$.
Assuming pure Hubble flow, we evaluate the pairwise separations between all voxels 
within the overdensity to find a maximum linear extent of $\Lmax \approx 19\,\hMpc$, 
which subtends over twice our effective spatial resolution element (defined as $2.3 \times$ our effective reconstruction
length of $3.5\,\hMpc$).
%% JFH What is our spatial resolution element. You never quote that anywhere do you? Also, if it is ~ 4 Mpc
%% the map smoothing, or comparable to the wiener kernel lenghts of 3 mpc/h, why is it only twice our resolution?
%% JFH put this senetence about the two minima right at the beginning
%% of paragraph when you describe the object you found.
Applying the `tomographic mass' estimation calibrated with our simulations (Equations~\ref{eq:fluxmass}, \ref{eq:mtomo}, 
and \ref{eq:mtomo_dm}), the total DM mass enclosed within the $\delsm/\sigsm<-3.5$ contour
is $M_{\rm dm} \equiv M_{\rm tomo} = \scien{(9 \pm 4)}{13}\,\hmsol$, where the uncertainty is estimated from the scatter
seen in the simulated protoclusters (Figure~\ref{fig:mtomo_mdm}).
%% JFH The M_dum and M_tomo should not be presented with tildas here. We have a relation and a calibration, and
%% you (should if you have not already) quote the error on this relation in the previous section. So these should
%% be measurements, not order of magnitue estimates which is what tilda signifies. We are much better than order
%% of magnitude acccuracy here.

%% JFH I think making Figure 12 and Figure 13 part of the same figure might make the readers life easier, i.e. flipping
%% back in forth between the two it is hard to see the corresondence. 

This IGM overdensity corresponds to a known object: D15 recently reported a $z=2.450$ protocluster
comprising of 11 LBGs, several of which we plot as filled squares in Figure~\ref{fig:pcslice1}.  
This protocluster was corroborated
by a more recent spectroscopic study of \lya\ emitters (LAEs) by C15, 
which are plotted as filled triangles in Figure~\ref{fig:pcslice1}. 
The position of these LBGs and LAEs at $z\approx 2.435 - 2.450$ appear
spatially well-correlated with our \lya\ overdensity at the same redshifts.
The tomography also shows that the elongated transverse configuration of the $z=2.450$ protocluster galaxies is due to 
two sub-structures at that redshift (Figure~\ref{fig:pcslice1}a) connected by a lower-density `bridge' 
well-sampled by our sightlines. 
Note that due to edge effects, it is possible that the protocluster 
might extend beyond our map boundary. 
Figures~\ref{fig:maplong} and \ref{fig:3d_pc} suggest that there could be a bridging filament between the $z=2.435$ and $z=2.450$ lobes
of the overdensity, although deeper observations with better sightline sampling and better spectral \snr\ would be required to verify this.

We now make an estimate of the descendant ($z=0$) mass of this protocluster based on its IGM tomography 
signature at $z=2.45$.
%% JFH We now make an estimate of the descendant (z=0) mass of this protocluster based its IGM
%% tomography signature at z=2.45
The $z=2.45$
tomographic mass enclosed within its $\delsm/\sigsm<-3.5$ boundaries implies, 
through Equation~\ref{eq:mtomo_m0}, that it will eventually collapse into a $M(z=0) = \scien{(3.1\pm 2.5)}{14}\,\hmsol$ 
object, i.e.\ an intermediate-mass Virgo-like cluster.
The uncertainty is, again, based on the scatter of in the simulated protoclusters (Figure~\ref{fig:mtomo_m0} 
with which we calibrated Equation~\ref{eq:mtomo_m0}.
%% JFH Mention the scatter here based on Figure 10?
The maximum smoothed absorption associated with the protocluster also provides an alternative estimate of the
$z\sim 0$ masses via Equation~\ref{eq:delmin_mass}. Applied to the absorption peak of 
$(\delsm/\sigsm)_\mathrm{min} \approx -4.2$ seen at $z=2.450$, 
this implies $z\sim 0$ masses of $M \approx \scien{(3.4\pm2.1)}{14}\,\hmsol$.
While these two estimates are not fully independent since $M_{\rm tomo}$ is calibrated through the 
map pixel values, it is reassuring that they give similar values.

\subsubsection{One or two $z=0$ clusters?}
From the unsmoothed tomographic map (Figure~\ref{fig:maplong} or \ref{fig:3d_pc}), 
the $z=2.45$ IGM overdensity is comprised of two separate lobes at $z\approx 2.435$ and 
$z\approx 2.450$. After imposing the smoothed $\delsm/\sigsm<-3.5$ cut for protocluster
selection, the resulting contour is continuous but still has two absorption peaks:
the more significant one ($\delsm/\sigsm = -4.2$) at $z=2.450$ with the other at
the $z=2.435$ lobe with $\delsm/\sigsm = -3.9$, with a comoving separation of $\approx 16\,\hMpc$ assuming
%% JFH Rather than ``no peculiar velocities'' say ``assuming only Hubble flow in the radial direction''
only Hubble flow in the radial direction.
Interestingly, if we split our evaluation of the underlying tomographic mass to either side of 
$z=2.442$ (the approximate saddle point between the two lobes) we find that the
the $z= 2.435$ part is actually slightly more massive at $M_{\rm tomo} \approx \scien{5}{13}\,\hmsol$
compared with $M_{\rm tomo} \approx \scien{4}{13}\,\hmsol$ at $z= 2.450$.
While this difference is insignificant compared to our uncertainties, it does point to the overdensity
being comprised of two roughly equal-mass portions at $\sim15\,\hMpc$ separation.
One might then wonder whether this protocluster candidate will in fact collapse into a single $z\sim 0$ 
cluster, or two separate ones. 
 
 \begin{figure}
 \includegraphics[width=0.49\textwidth]{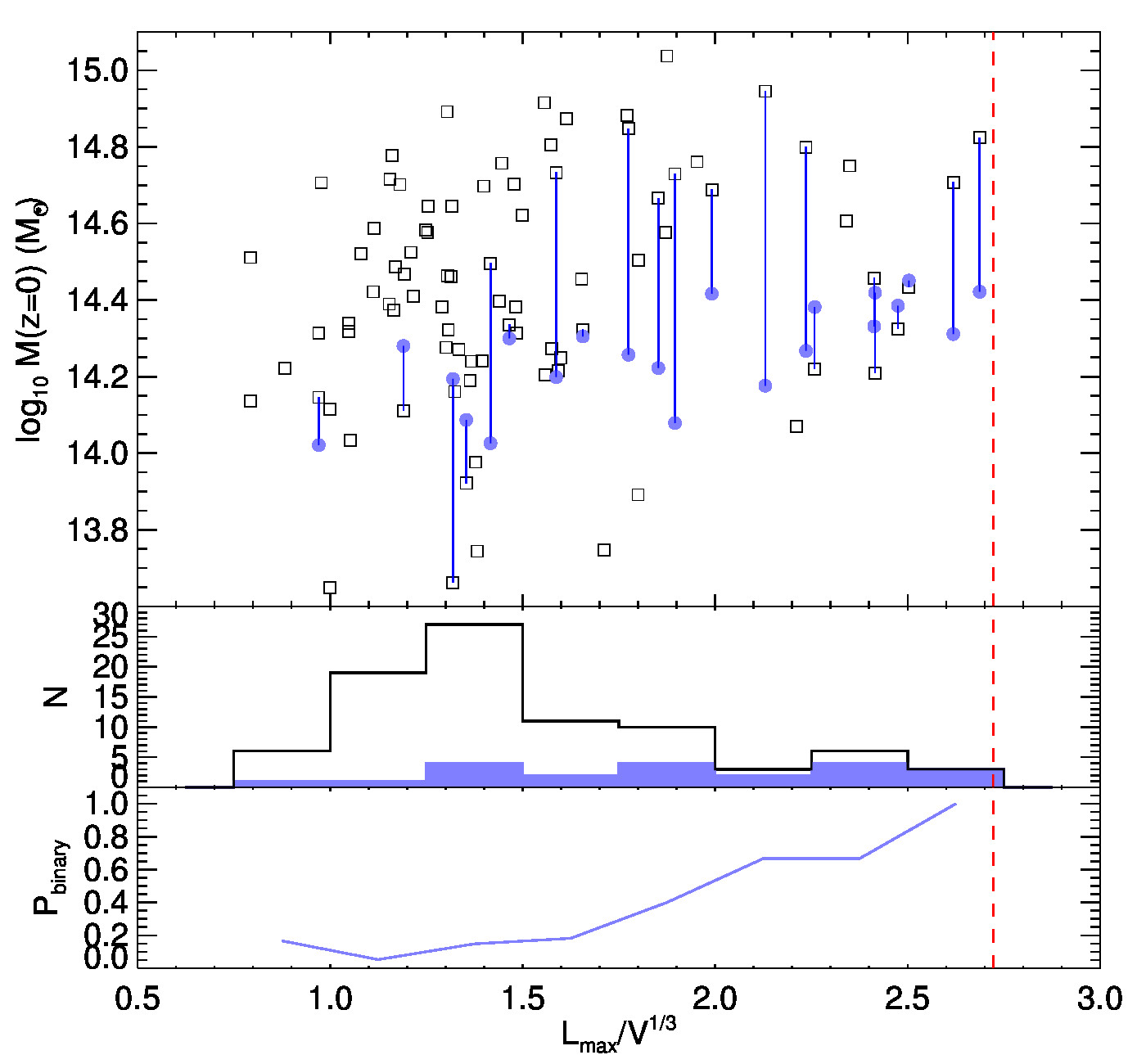}
 \caption{\label{fig:elonmass}
 The elongation of IGM protocluster candidates identified in the simulated tomographic maps. In the top panel, 
 we show the descendant mass $M(z=0)$ associated with each protocluster candidate as a function of 
 the elongation ratio $\Lmax/V^{1/3}$. In the cases where there is a second cluster progenitor associated
 with the IGM candidate, the secondary protocluster is indicated with a blue circle and connected to the primary
 with a blue line. The middle panel shows the distribution of $\Lmax/V^{1/3}$ with the subset of binaries in blue, 
 while the bottom panel shows the binary fraction at each $\Lmax/V^{1/3}$ bin. Vertical red dashed line indicates
 the value from the $z=2.45$ protocluster in our observed map.
 }
 \end{figure}

To provide some insight, we again turn to the simulated protoclusters
detected in tomography. S15 described such
%% JFH bad merge sounds like it is an artifact of linking procedure. I would omit the name in the text and maybe put
%% that in a footnote instead, i.e. ``These were referred to as bad-merge object by Stark, although they correspond
%% to dinstince M> 10^14 halos at z =0 or something. 
situations in the protocluster search process, 
in which a single $z=2.5$ IGM protocluster candidate is in fact associated with progenitors from two 
separate $z=0$ clusters\footnote{These were referred to as ``bad-merge'' objects by S15, 
although they correspond to distinct $M> 10^{14}$ halos at $z =0$}. 
Out of the 81 protoclusters identified through our simulated $(256\,\hMpc)^3$ tomographic map, 
we find that 21 ($\sim 25\%$) could in fact be associated with a secondary $z=0$ cluster.
We noticed that many of these `binary' protoclusters appeared elongated (e.g.\ middle panel in Figure~\ref{fig:protoreal}),
so we compute $\Lmax/V^{1/3}$ on their $\delsm/\sigsm<-3.5$ boundaries 
as a rough measure of their elongation. This is shown in Figure~\ref{fig:elonmass}, which compares
$\Lmax/V^{1/3}$ with $M(z=0)$ for both unitary and binary protoclusters.
In the lower panels of Figure~\ref{fig:elonmass}, we see that the binary fraction of the IGM protocluster
candidates is low ($\sim 10\%$) for roughly spherical ($\Lmax/V^{1/3} \sim 1$) overdensities
but increases with $\Lmax/V^{1/3}$ and approaches unity at $\Lmax/V^{1/3}\sim 2.5$, i.e.\  
protocluster candidates that appear highly elongated in the smoothed IGM tomographic maps are
likely to evolve into two separate clusters by $z\sim 0$.

Our $z= 2.45$ IGM overdensity has a maximum extent of $\Lmax \approx 19\,\hMpc$ and an 
enclosed volume of $V\approx 340\,h^{-3}\,\mathrm{Mpc}^3$, which leads to $\Lmax/V^{1/3} \sim 2.7$.
This extreme value is only sampled by a handful of simulated protocluster candidates in Figure~\ref{fig:elonmass}, 
but we see that $\sim 80\%$ of simulated protoclusters with $L/V^{1/3} > 2$ collapsed into two separate
$z=0$ clusters.
This indicates a high probability that the IGM overdensity in our map will eventually
collapse into two separate $z=0$ clusters with $M(z=0)\sim 10^{14}\,\hmsol$, 
centered on the two absorption peaks at $z= 2.435$ and $z= 2.450$.
%% JFH Be quantitative about that probability, i.e. XX% of simulated proto-clusters with L/V^1/3 > YY collapsed into
%% two clusters at z =0. 

\subsubsection{Less Significant IGM Overdensities}
Apart from the significant $z=2.450$ IGM overdensity in our map that fulfils the protocluster criterion defined by S15, 
there are also several smaller `hot-spots'  that can be seen in the unsmoothed map
(Figures~\ref{fig:3d_pano} and \ref{fig:maplong}), none of which fulfill the $\delsm<-3.5\,\sigsm$ threshold
due to their limited size. 
In Slices \#4-6, there are two smaller overdensities at $z\approx2.23$ and $z\approx2.25$
with transverse extents of $\lesssim5\,\hMpc$ in the lower part of the map 
($\yperp \approx 6\,\hMpc$ and $\yperp\approx 3\,\hMpc$, respectively). 
These overdensities have map values of $\delsm\approx [-2.1, -1.8]\,\sigsm$ after smoothing
and thus do not satisfy our protocluster criterion, although there are several galaxies associated with a weaker
extension of the $z\approx2.23$ structure in Slices \#5 and \#6.
%% JFH Now you are confusing what we call significant. The original cluster-finding is done using the \sigma_f*3.5
%% criterion which includes in all sources of noise. Now you are adopting by eye sampling criterion which are confusing
%% and have not been discussed before. Why not just say these are 2.8sigma and 2.0 sigma according to that criterion,
%% and hence are not significant. It is not clear to me that the isnigificance here is actually a sampling issue. 
There are another two apparent overdensities in Slice \#1 at $z\approx2.27$ and $z\approx2.48$ at $\yperp\approx 3\,\hMpc$
and $\yperp \approx 9\,\hMpc$, respectively.
%% JFH Always quote both the y_perp value and the redshift, so people can see what you are talking about. 
Neither of these overdensities ($\delsm \sim [-1.4, -1.8]\,\sigsm$) satisfy our protocluster
criterion, although this part of the map likely suffers from edge effects. A better characterization of these possible overdensities
would require an extention of the map area beyond its current boundaries.
%% JFH Also quote the sigma signifciacnes of these two things. 

\begin{figure*}
\begin{center}
\includegraphics[height=0.04\textheight]{colorbar_pcslice.jpg}\\  
\begin{overpic}[width=0.43\textwidth,clip=true,trim=0 24 0 0]{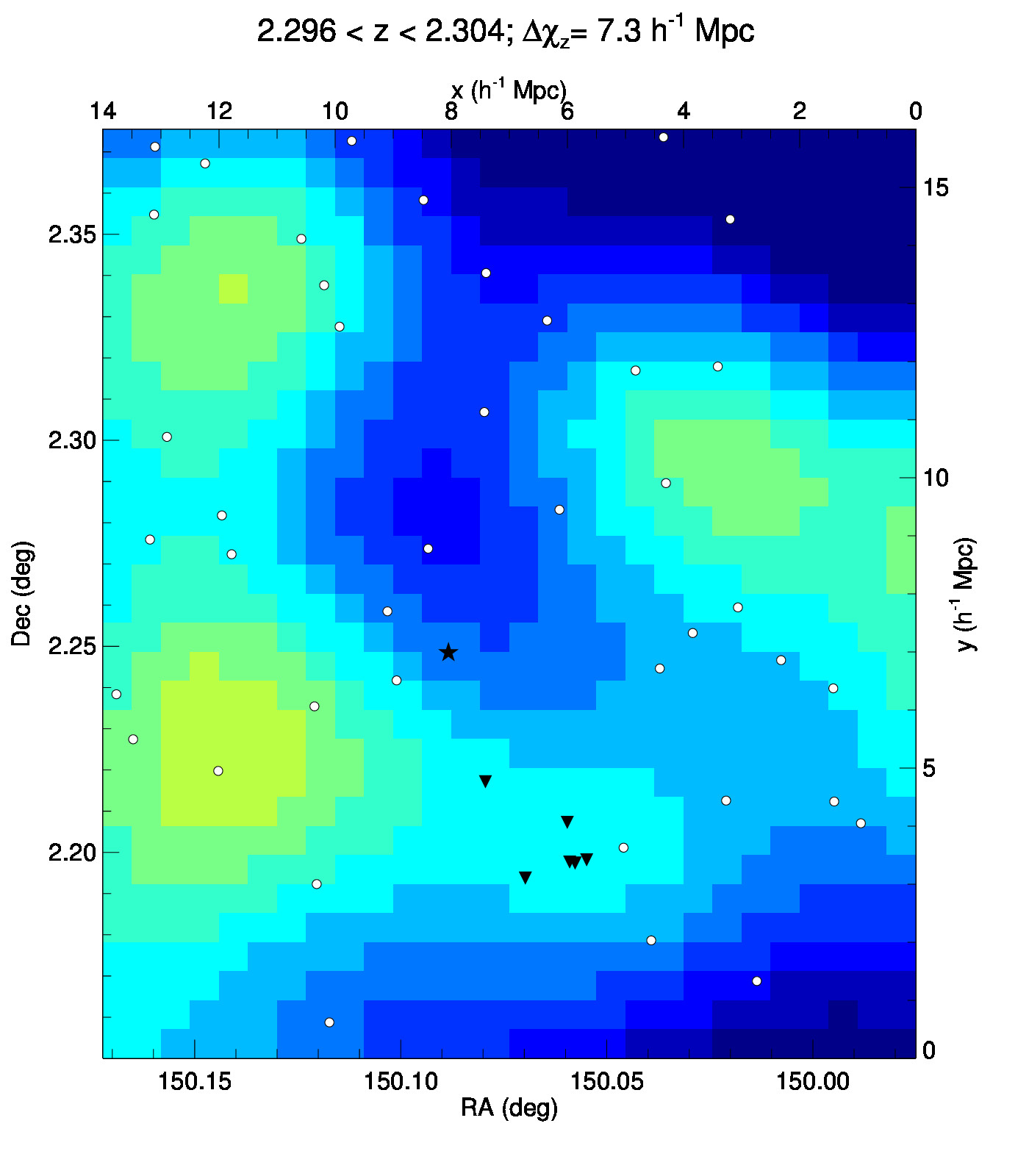}
\put(42,233){\textbf{(a)}}
\end{overpic}
\begin{overpic}[width=0.43\textwidth,clip=true,trim=0 24 0 0]{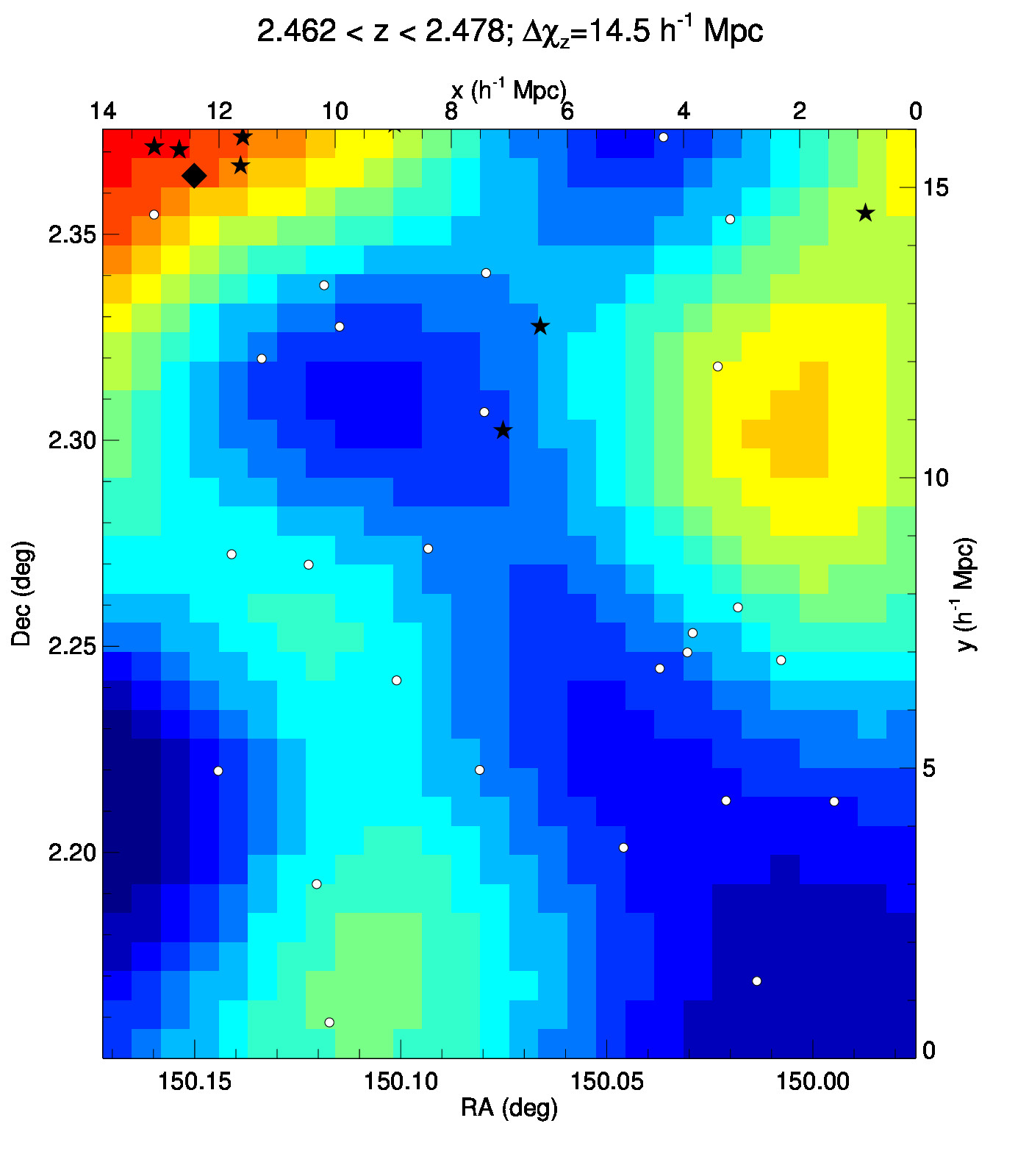}
\put(42,233){\textbf{(b)}}
\end{overpic}
\end{center}
\caption{\label{fig:pcslice2}
Same as Fig.~\ref{fig:pcslice1}, but focused on two galaxy overdensities within our map.
Panel (a) is focused on a compact overdensity of MOSDEF galaxies (upside-down triangles),
while (b) shows our partial coverage of the $z\approx 2.47$ protocluster
recently reported by \citet{casey:2015}. 
Open circles indicate our sightline sampling, while filled symbols indicate coeval galaxies within the map: 
MOSDEF galaxies (inverted triangles, K15); zCOSMOS-Deep LBGs \citep[stars,][]{lilly:2007}, and a
DSFG \citep[diamond,][]{casey:2015}.}\end{figure*}

\subsection{Galaxy Overdensities}\label{sec:galoverden}
Apart from the overdensity in the \lya\ absorption map, we now discuss two galaxy overdensities
from external data sets.
\subsubsection{$z=2.30$}
During our comparison of the map overdensities and coeval galaxies, we noticed a compact overdensity of MOSDEF
galaxies at $z\approx2.30$ centered on [RA, Dec] = [150.06, 2.20]deg (pink circles in Slices \#3-4 of 
Figure~\ref{fig:maplong});
%% JFH When referring to Figure 4, refer to slice numbers
the line-of-sight projection centered on this redshift is also shown in \ref{fig:pcslice2}a). 
This is comprised of 6 galaxies within $\approx2\,\hMpc$ 
(or $\approx1\,$pMpc) on the sky, and a line-of-sight extent of $\Delta z\approx0.006$ or $\Delta v\approx560\,\kms$, 
corresponding to $\approx 6\,\hMpc$ ($\approx 2.6$pMpc) assuming Hubble flow.
If these galaxies fairly trace the underlying population, then within the subtended comoving volume the number density is
$\rho_\mathrm{gal}\sim 0.3\,h^3\, \mathrm{Mpc}^{-3}$, which translates to an overdensity of 
$\rho_\mathrm{gal}/\langle \rho_\mathrm{gal}\rangle \sim 40$ relative to a mean
number density of 
$\langle \rho_\mathrm{gal} \rangle \sim \scien{7}{-3}\,h^3\,\mathrm{Mpc}^{-3}$, estimated from
 $2<z<3$ sources with $H\leq 24.5$ in the \citet{santini:2015} 
CANDELS/GOOD-S catalog (see \citealt{lee_ks:2013} for a similarly compact $z\sim 3.7$ overdensity). 
Within this structure, the quartet of galaxies at [RA, Dec] $\approx [150.058, 2.200]$deg
is even more compact: with a sky footprint of $\sim 0.17\mathrm{arcmin}^2$ and $\Delta z \approx 0.003$, 
this implies 
$\rho_\mathrm{gal}/\langle \rho_\mathrm{gal}\rangle \gtrsim 100$.

%% JFH transverse comoving radius is confusing. Just say within a radius of 1 Mpc/h on the sky, and (perhaps
%% giving a central RA and DEC, and then for the line-of-sight velocity, quote what this corresponds to in
%% terms of LOS distance assuming Hubble flow, i.e. about 6 Mpc/h. 

%% I don't understand what you mean by ``when ranked by increasing
%% overdensity''. This is confusing, because we measure over-flux and not
%% actually overdensity. Just say 65% of the cumulative distribution
%% P( <\delta_F) of \delta_F.
In Figures~\ref{fig:maplong} and \ref{fig:pcslice2}a, we do not see any large 
absorption decrement associated with these galaxies, although the sightline sampling
could be better.
%% JFH --> could be better
These galaxies occupy map values of $\delrecon\sim-0.1$ or $\delsm \sim -0.02$
after smoothing, therefore based on the calibration between smoothed flux and DM overdensity (Equation~\ref{eq:fluxmass}),
this suggests, at face value, that the MOSDEF galaxies occupy a region
with $\rho/\bar{\rho}\sim1$, i.e.\ the mean density of the Universe.
%% JFH Mention or quote the 1-sigma scatter in this relation, i.e. the scatter in Figure 8. 
This is consistent with the ZFOURGE distribution of $K<24.8$ galaxies with photometric 
redshifts $2.1<z<2.3$ \citep{spitler:2012} in the same field, which appears to show roughly mean density at the 
position of these MOSDEF galaxies, 
although $z=2.300$ is at the edge of the redshift range in the ZFOURGE map. Note that since their map is 
smoothed over a much larger radial window ($\Delta z =0.2$) than the $\Delta z \approx 0.006$ line-of-sight extent of 
the MOSDEF overdensity, the ZFOURGE map is not in conflict with the presence of the MOSDEF overdensity.

In Figure~\ref{fig:threshist}, we indicate the $\delsm/\sigsm$ value associated with this
overdensity after smoothing with a $\sigma=4\,\hMpc$ Gaussian, i.e. the quantity we used to search for protoclusters.
This corresponds to $\delsm/\sigsm \approx -0.4$, which is far from our $\delsm/\sigsm <-3.5$ protocluster threshold
and at a value characteristic of typical regions of the map.
%% JFH and at a value characteristic of typical regions of the map. 
In comparison with the simulations we see that none of the 425 protoclusters within the volume
have such a weak absorption despite very diverse samplings of sightline and spectral \snr\ across these
protoclusters. In other words, even the worst-sampled protoclusters within our simulated tomographic
reconstructions have stronger large-scale IGM absorption than this overdensity of galaxies.
Furthermore, as can be seen in the bottom panel of Figure~\ref{fig:protoreal}, even protoclusters
that do not satisfy our $\delsm<-3.5\,\sigsm$ IGM detection threshold
still exhibit moderate absorption over large ($\sim 10\,\hMpc$) scales which we do not
see in the vicinity of these MOSDEF galaxies.

One possibility is that galaxy formation feedback has suppressed IGM opacity in the immediate
vicinity of these galaxies on scales of our $\sim 3.5\,\hMpc$ effective smoothing of our tomographic
reconstruction. 
This could a similar effect observed by \citet{adelberger:2005}, 
who claimed a detection of elevated \lya\ transmission within $\sim 0.5\,\hMpc$ of 
$z\sim3$ LBGs due to galactic winds, although this effect was not seen in hydrodynamical simulations 
\citep{kollmeier:2006, viel:2013b},
nor did subsequent observations \citep{steidel:2010,crighton:2011} manage to reproduce this result.
%% JFH Next to Crighton also cite the updated paper by Kurt where this effect largely went away. Just to be polite. 
However, due to the compactness (mean 3D comoving separation $\sim 0.6\,\hMpc$) of this $z=2.30$ galaxy overdensity it is possible that 
the effect of feedback has been enhanced beyond that observable in individual LBGs.
Another possibility is that a hot ($T>10^6\,\mathrm{K}$)
%% JFH Well even at the group scale, virial temperature is 10^7, so I'd make this T > 10^6, or T ~ 10^6-7
intra-cluster medium (ICM) has already 
formed from the virialization of such a compact
overdensity, similar to that observed in X-rays for a $z=2.07$ cluster \citep{gobat:2011}. 
The presence of a hot ICM might
%% JFH would --> might
also suppress the
\lya\ forest absorption on scales of $\sim 1\,\hMpc$ (assuming a virial mass of $M\sim 10^{13.5}\,\hmsol$).

It is possible that a better sampling could reveal a stronger overdensity in the immediate vicinity 
(within $\lesssim 1-2$Mpc)
of this galaxy overdensity, but there is little strong absorption in the neighboring sightlines we do have, 
which constrains the transverse extent of  
the associated IGM absorption to $<4\,\hMpc$ comoving, or $<2\,$Mpc physical.
This lack of large-scale overdensity suggests that this association of galaxies is
unlikely to grow into a massive galaxy cluster by $z\sim0$ \citep{chiang:2013}
due to the lack of available material to accrete from its large-scale environment.
We demonstrate this by applying a very conservative toy galactic feedback model on our simulated tomographic maps:
for all map voxels within $1.5\,\hMpc$ (approximately the virial radius of $\sim 2\times10^{14}\,\hmsol$ halo) of the
absorption peak associated with each protocluster,
%% JFH Need to make some argument like this 1.5 Mpc/h is about the virial radius of a mass M ~ 10^XX virialized
%% collapsed halo
%% that you are imagining might host these objects. 
we set
$F=1$ or $\delta_F=0.25$ (since we adopted $\langle F\rangle=0.8$), i.e. zero \lya\ absorption, and then smooth the maps 
by a $\sigma=4\,\hMpc$ Gaussian kernel and evaluate the minimum $\delsm/\sigsm$ within $4\,\hMpc$ of
each protocluster center as before. The resulting distribution is shown as the red dot-dashed histogram in Figure~\ref{fig:threshist},
which is shifted towards higher $\delsm/\sigsm$ compared to the original protocluster distribution as expected, 
but only by a small amount insufficient to match the low map absorption associated with the $z=2.300$ MOSDEF overdensity.

%% JFH Rather than continually re-stating this sampling noise caveat, why don't you just say something quantitative,
%% like: We considered X different sampling and noise realizations of our simulated clusters, and found that in
%% 0 cases out of XXX did a real > 10^14 Msun (z=0) proto-cluster scatter to a level as low as the MOSDEF object value. 
%% This averages over all possible sampling configurations. Then quote an upper limit on the probability as being
%% the Poisson upper limit associated by zero divided by XXX. 
While it is possible that we have been unlucky with our sightline sampling and there is in fact an extended IGM overdensity
that just happened to avoid our sightlines, the current data implies that regardless of any possible effects from 
galactic feedback or a hot ICM on scales of $\lesssim 1\,\hMpc$, 
this $z=2.300$ overdensity of galaxies appears unlikely to evolve into a $M\gtrsim 10^{14}\,\hmsol$ galaxy cluster by $z\sim 0$, 
and could instead be a precursor of a compact galaxy group \citep{hickson:1997}.  
This contrasts with the similarly compact $z=3.8$ LBG overdensity found by \citet{lee_ks:2013}, in which follow-up
observations revealed a large-scale LAE overdensity \citep{lee_ks:2014}.

%% JFH I would visually inspect the spectra that contributed to the
%% MOSDEF non-cluster, make sure they are not fucked up in some way,
%% and then add a footnote to the paragraph above saying ``visual
%% inspection of the individual spectra contributing to the map at
%% this location confirms that there is not significant IGM absorption
%% in any of the neighboring sightlines. 

\subsubsection{$z=2.48$}
Recently, \citet{casey:2015} reported a protocluster that was initially identified as a close association
of dusty star-forming galaxies (DSFGs), but was also shown to correspond to an LBG overdensity.
While the reported center of this overdensity lies outside our map volume, several of the protocluster members
fall just within the edge of our map boundary. 
Their projected transverse positions are juxtaposed with our map in Figure~\ref{fig:pcslice2}b. Encouragingly, 
there is an IGM overdensity associated with these galaxies 
right at the edge of our map,  which is sampled by two sightlines: one in the top-left corner of our map area 
(Figure~\ref{fig:pcslice2}b) and another just outside the map boundary (sightline 15175 in Figure~\ref{fig:targets}).
Clearly, more data would be required to carry out a detailed investigation of this structure.
%% JFH If more than just the one spectrum hat is shown on the overdensity contributed to the map, i.e. because
%% you used sources outside the map volume, you should add a note or footnote saying that here. Otherwise it looks
%% just like we had that single sightline. 

%% JFH The two different sets of units that are you using for Dec in Figure 5 and in Figure 2 are in my opinion
%% confusing. I actually think sexigecimal is most appropriate here, so just stick with that. 

\section{Conclusion}
In this paper, we describe \lyaf\ tomographic reconstructions of the $2.2<z<2.5$ \lyaf\ from
58 background LBG and QSO spectra within a
$\sim12\arcmin\times14\arcmin$ area of the COSMOS field, which maps the IGM large-scale structure 
with an effective smoothing scale of $\sim3.5\,\hMpc$ over a comoving volume of $V\approx5.8\times10^{4}\,\h^{-3}\mathrm{Mpc}^3$.

Our main findings can be summarized as follows:
\begin{itemize}\item We compared our map with 61 coeval galaxies with known spectroscopic 
redshifts from other surveys. These galaxies preferentially occupy highly-absorbed regions of the map at high statistical
significance;
this skew is even stronger with the subsample of 31 galaxy redshifts measured with NIR nebular lines 
from MOSDEF (K15), which suffer less line-of-sight positional error that could scatter galaxies 
into underdense map regions. Future tomographic maps encompassing hundreds of coeval galaxies will allow an 
investigation of $z\sim 2-3$
%% JFH Make this z \sim 2-3?? to sound better. 
galaxies as a function of their large-scale IGM environment.
\item After applying the smoothed flux threshold advocated by S15 for finding
galaxy protoclusters on \lyaf\ tomographic maps, we find one significant and extended 
($\Lmax \approx 19 \,\hMpc$) IGM overdensity at 
$z\approx 2.44$ that is associated
with a galaxy protocluster reported by \citet{diener:2015} and \citet{chiang:2015}. 
Within the $\sim 340\,h^{-3}\,\mathrm{Mpc}^3$ volume of this overdensity, we estimate an enclosed
mass of $M_{\rm dm}(z=2.5) =\scien{(9\pm4) }{13}\,\hmsol$.
%% JFH again, equal sign with an error. This is a measurement!!
In comparison with the distributions of size and IGM absorption depth of simulated protoclusters,
we argue that this object
will likely collapse into a $M(z=0)\approx 3\times 10^{14}\,\hmsol$ cluster, although from its highly 
elongated morphology ($\Lmax/V^{1/3} \sim 2.7$) there is  
a high probability that it will collapse into two separate clusters by $z\sim0$.
Deeper IGM tomographic observations to fully characterize the morphology of this overdensity, 
along with detailed modeling incorporating the associated
member galaxies, could allow us better distinguish between these scenarios.

\item Within our map volume, we note a compact overdensity of six MOSDEF galaxies at $z=2.300$ within a $\sim 1\,\hMpc$ transverse
radius and $\sim 5\,\hMpc$ along the line-of-sight. There is no large IGM overdensity associated with 
these galaxies; rather, they occupy a region of approximately mean absorption on scales of several Mpc. 
While it is possible that galaxy feedback or a hot ICM has suppressed the \lya\ absorption in the immediate vicinity ($\lesssim 1\,\hMpc$)
of this overdensity, the lack of an extended $r\gtrsim 5\,\hMpc$ overdensity
 implies that they are unlikely to grow into a massive cluster by $z\sim0$.
However, the current sightline configuration does not sample this region well, and 
more observations will be needed to better characterize the environment of this overdensity. \end{itemize}
%% JFH I think you should kill this caveat here, and just do the probability estimate I mention above 

These observations were from the pilot phase of the
upcoming CLAMATO survey, which is aimed at mapping the $z\sim2-3$
%% JFH z ~ 2-3 sounds better
IGM across a $\sim 1\,\sqdeg$ area
of the COSMOS field. This pilot study
%% JFH This pilot study validates
validates the ability of \lyaf\ tomographic reconstructions to study  
large-scale structure at these unprecedentedly high redshifts, 
particularly protocluster overdensities that will eventually collapse into massive
galaxy clusters by $z\sim 0$. 
%% JFH Take out these weather allocation caveats, and just a 1 deg^2
%% mapped area would give this volume and these clusters.
A $1\,\sqdeg$ survey over $2.2<z<2.5$ would yield comoving volume of $V\sim 10^6\,\hMpc$ would yield $\sim 5$ protocluster detections with $M(z=0)>10^{14}\,\hmsol$
of which $\sim 3$ would be progenitors of Virgo-like clusters ($M(z=0) \gtrsim 3\times10^{14}\,\hmsol$ or larger; see also S15).
%% JFH This last sentence needs to have the word proto-cluster in it. 

Apart from identifying protoclusters with blind tomographic surveys such as CLAMATO, 
the insights from this paper suggest that it would also be profitable to carry out similar observations
 targeted at known or suspected protoclusters at $z\sim 2-3$.
Indeed, over such limited fields it could be worthwhile to pursue higher-\snr\ integrations on 
higher area-densities of background sources than we have achieved here, which would
allow reconstructions on smaller scales with reduced map noise \citep[see][for a detailed discussion]{lee:2014}. 
In synergy with
%% JFH spectroscopic confirmation of 
spectroscopic confirmation of coeval member galaxies and
%% JFH bolsted by detailed modeling with simulations like those used here
boosted by detailed modeling with simulations like those used here, such observations could reveal the likely
assembly mechanism of individual protoclusters, which would allow workers to build up a global picture of massive
cluster formation from high-redshifts to $z\sim 0$.

%% JFH I think you should add a few sentences in the conclusions also plugging and prospecting for the voids.

\bibliographystyle{apj}

\acknowledgements{We are grateful to the entire COSMOS collaboration for their assistance and helpful discussions.
 J.F.H. acknowledges generous support from the Alexander von Humboldt foundation in the context of the Sofja Kovalevskaja Award. The Humboldt foundation is funded by the German Federal Ministry for Education and Research.
  The authors also wish to recognize and acknowledge the very significant cultural role and reverence that the summit of Maunakea has always had within the indigenous Hawai'ian community.  We are most fortunate to have the opportunity to conduct observations from this mountain.}
%% JFH Please add my boiler plate acknowledgement:
% J.F.H. acknowledges generous support from the Alexander von Humboldt foundation in the context of the Sofja Kovalevskaja Award. The Humboldt foundation is funded by the German Federal Ministry for Education and Research. 

\bibliography{lyaf_kg,apj-jour,lss_galaxies,my_papers}

\end{document}